\documentclass[%
 reprint,
 superscriptaddress,
preprintnumbers,
nofootinbib,
 amsmath,amssymb,
 aps,
prl,
]{revtex4-1}
\pdfoutput=1

\usepackage{amsmath}
\usepackage{amssymb}
\usepackage{amsthm}
\usepackage{MnSymbol}
\usepackage{nicefrac}
\usepackage{amsfonts}
\usepackage{dsfont}

\usepackage[markup=nocolor]{changes}
\usepackage{natbib}

\usepackage{graphicx}

\usepackage[normalem]{ulem}

\usepackage{dcolumn}
\usepackage{bm}

\usepackage{color}
\usepackage{xcolor}
\newcommand{\bea}{\begin{eqnarray}}
\newcommand{\eea}{\end{eqnarray}}
\newcommand{\be}{\begin{equation}}
\newcommand{\ee}{\end{equation}}

\begin{document}

\title{Image features of spinning regular black holes based on a locality principle}
 
 \author{Astrid Eichhorn}
   \email{eichhorn@cp3.sdu.dk}
\affiliation{CP3-Origins, University of Southern Denmark, Campusvej 55, DK-5230 Odense M, Denmark}
\author{Aaron Held}
\email{a.held@imperial.ac.uk}
\affiliation{Theory Group, Blackett Laboratory, Imperial College London, SW7 2AZ, London, UK}

\begin{abstract}
To understand the true nature of black holes, fundamental theoretical developments should be linked all the way to observational features of  black holes in their natural astrophysical environments. Here, we take several steps to establish such a link. We construct a family of spinning, regular black-hole spacetimes based on a locality principle for new physics and analyze their shadow images. We identify characteristic image features associated to regularity (increased compactness and relative stretching) and to the locality principle (cusps and asymmetry) that persist in the presence of a simple analytical disk model.
We conjecture that these occur as universal features of distinct classes of regular black holes based on different sets of construction principles for the corresponding spacetimes.
\end{abstract}

\pacs{Valid PACS appear here}

\maketitle
General Relativity (GR) has passed a multitude of observational tests~\cite{Will:2014kxa,Berti:2015itd}. In particular, it describes spacetime outside the event horizon of a black hole consistently with observations within the statistical and systematic uncertainties~\cite{TheLIGOScientific:2016src,Hees:2017aal,LIGOScientific:2019fpa,paper6,Abbott:2020jks}. Despite these successes, a glaring problem with the theoretical description of black holes in GR remains. The theory predicts a singularity of the space-time curvature, signalling its own breakdown. Therefore, we do not understand the true nature of black holes. We know from observations that compact objects very much like black holes exist~\cite{Cardoso:2019rvt}. We know from theoretical consistency that these compact objects must be nonsingular. Thus, a two-step development is called for, in which theoretically consistent, regular spinning black-hole spacetimes are constructed with a subsequent extraction of image features relevant for the Event Horizon Telescope~\cite{paper1, paper2, paper3, paper4, paper5, paper6} as well as observables accessible to other observations, e.g.,~\cite{Flachi:2012nv,Carballo-Rubio:2019fnb,Panotopoulos:2019qjk,Zhou:2020eth,Liu:2020ola,Agullo:2020hxe}.\\
Here, we propose a family of space-time metrics for spinning, regular black holes based on a locality principle. We derive observational signatures in their shadows that are characteristic imprints of the locality principle. Specifically, our analysis accounts for observationally relevant parameters like spin and inclination and includes a simple analytical disk model following~\cite{2020ApJ...897..148G}.

In order to take key steps along the way from fundamental theory to future observational data we i) start from a fundamental theoretical principle, ii) construct a family of spacetime metrics based on it, and iii) calculate images in a toy model for the astrophysical environment of the black hole.
\\

\emph{Four principles for black-hole spacetimes}:
In contrast to parameterized deviations from the Kerr spacetime, see, e.g.,~\cite{Vigeland:2009pr,Johannsen:2011dh,Vigeland:2011ji,Cardoso:2014rha,Johannsen:2015pca,Lin:2015oan,Konoplya:2016jvv}, we base our construction on four fundamental physical principles. We demand that
\begin{enumerate}
\item[i)] the spacetime is described by a metric that is a solution to some dynamics beyond GR,
\item[ii)] the metric has a correct Newtonian limit,
\item[iii)] the spacetime is nonsingular everywhere,
\item[iv)] deviations from Kerr follow a locality principle encoded in the local value of spacetime curvature.
\end{enumerate}
The last principle is what distinguishes our construction from other works in the literature, e.g.,~\cite{Reuter:2010xb,Bambi:2013ufa,Litim:2013gga,Azreg-Ainou:2014pra,Toshmatov:2014nya,Ghosh:2014hea,Pawlowski:2018swz,Kumar:2019ohr,Contreras:2019cmf,Liu:2020ola,Junior:2021atr,Mazza:2021rgq}. As we will see, its implementation results in distinct new spacetime and image features.
It follows from a (classical or quantum) effective field theory point of view on new physics. This point of view implies that deviations from GR set in beyond a critical value of the local curvature. It is motivated by several research lines in the literature, where both quantum as well as classical modifications of GR follow effective field theory principles, both for black holes, e.g.,~\cite{BjerrumBohr:2002ks,Lu:2015cqa,Cardoso:2018ptl,Konoplya:2020bxa,Borissova:2020knn,deRham:2020ejn,Xie:2021bur} and beyond, e.g.,~\cite{Starobinsky:1980te,Avramidi:1985ki,Donoghue:1994dn,Sotiriou:2008rp,DeFelice:2010aj,Alvarez-Gaume:2015rwa,Levi:2018nxp,Eichhorn:2018yfc,Kobayashi:2019hrl}.
Within the highly restricted setting of spherical symmetry, principles (i)-(iv) have been successfully implemented, cf.~\cite{Dymnikova:1992ux,AyonBeato:1998ub,Hayward:2005gi,Platania:2019kyx,Simpson:2019mud}. However, astrophysical black holes typically spin~\cite{Gammie:2003qi,Reynolds:2013qqa,Abbott:2020gyp,2020arXiv201108948R}. 
Yet, to the best of our knowledge, no example in the literature implements principle iv) for a spinning black hole. This is rooted in the intricate spacetime and curvature structures of Kerr black holes: for instance, the Kretschmann scalar changes its sign at various locations throughout the spacetime. This makes it technically challenging to define an invariant local curvature scale as a new-physics scale.
In fact, explicit constructions of regular spinning black holes to date only implement principles i) -iii)~\cite{Reuter:2010xb,Bambi:2013ufa,Litim:2013gga,Azreg-Ainou:2014pra,Toshmatov:2014nya,Ghosh:2014hea,Pawlowski:2018swz,Kumar:2019ohr,Contreras:2019cmf,Liu:2020ola,Junior:2021atr,Mazza:2021rgq}.

\begin{figure*}
	\begin{center}
		\begin{minipage}{0.94\linewidth}
		\includegraphics[height=0.33\linewidth]{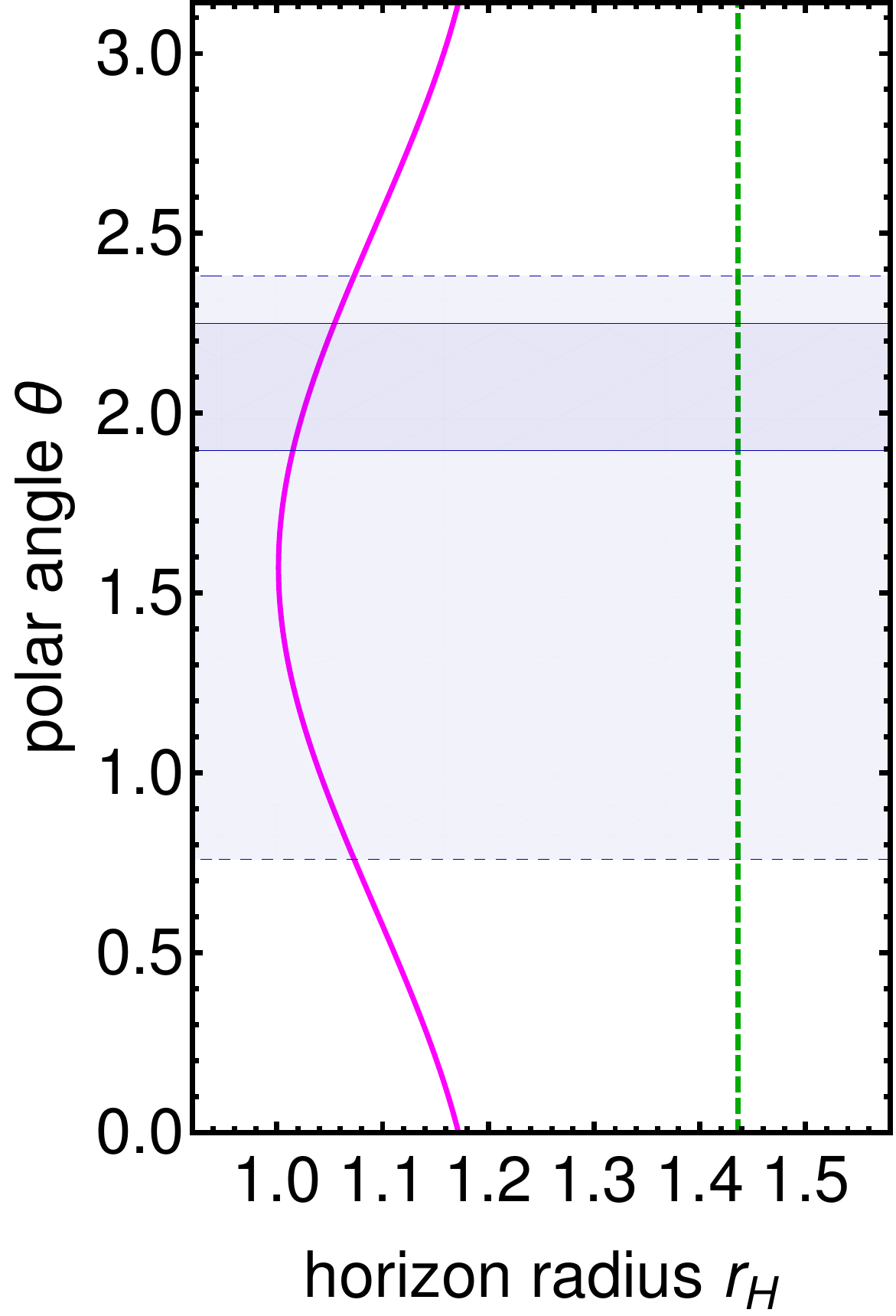}\hfill
		\includegraphics[height=0.33\linewidth]{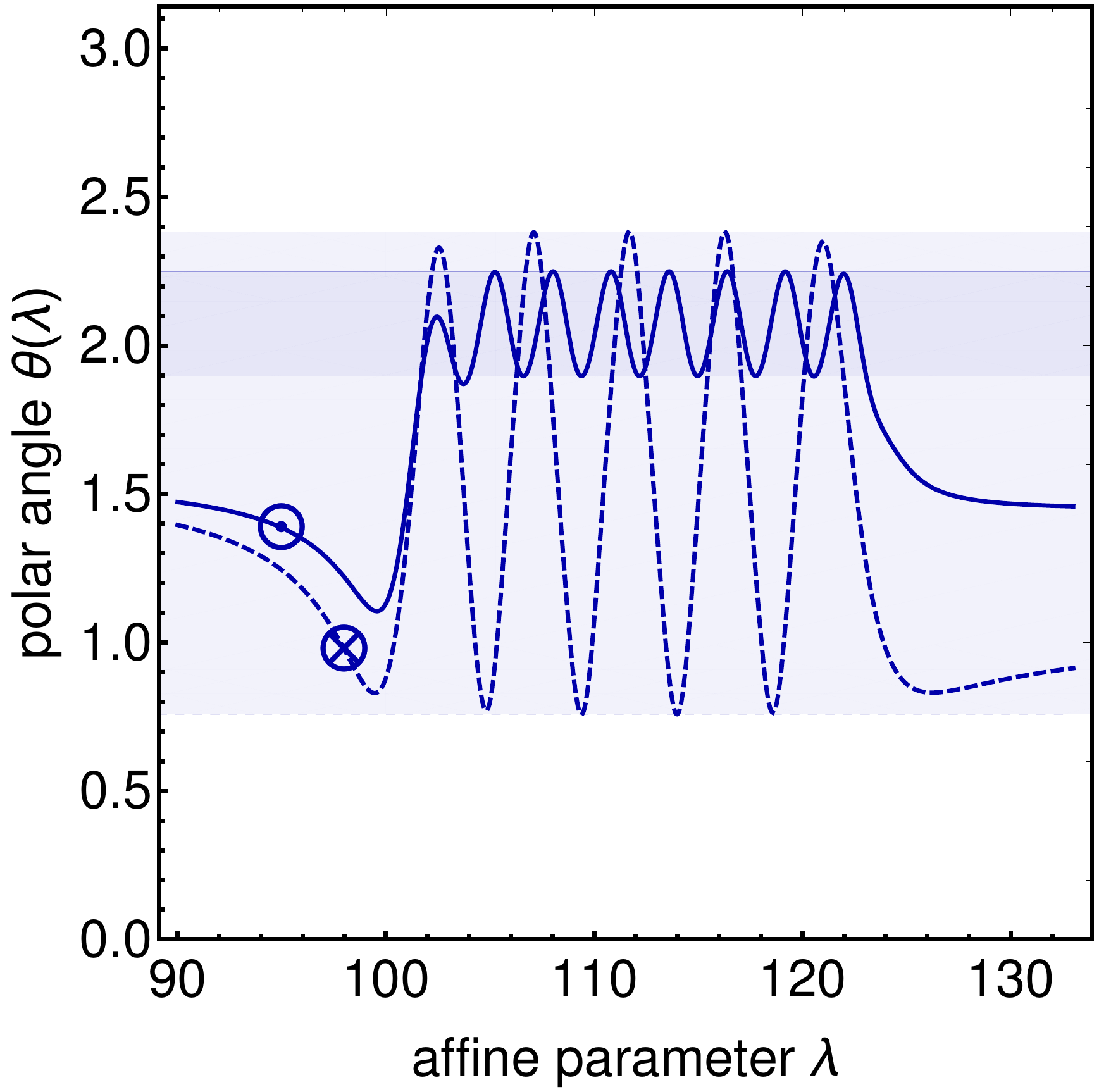}\hfill
		\includegraphics[height=0.33\linewidth]{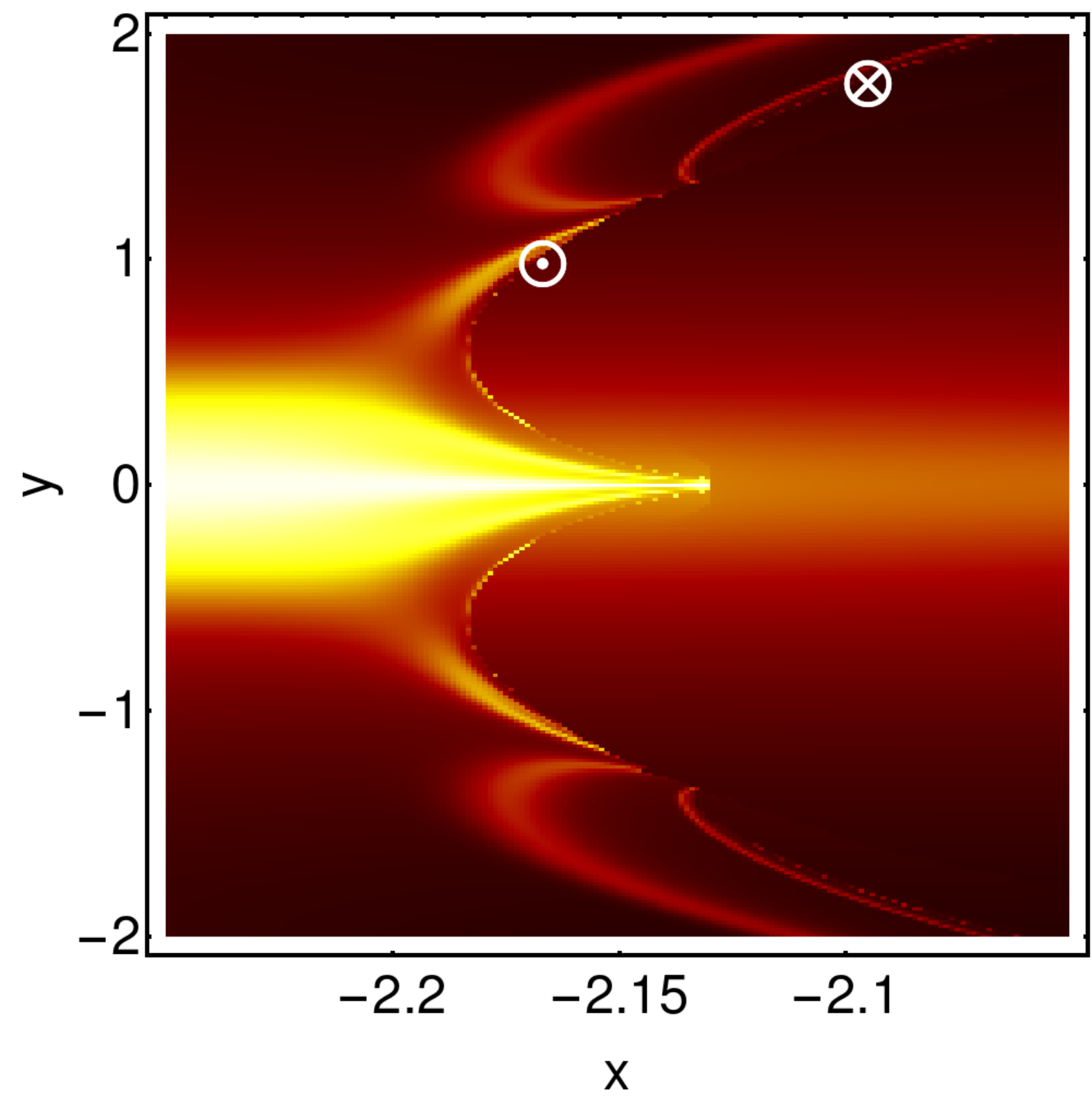}\hfill
		\end{minipage}
		\begin{minipage}{0.05\linewidth}
			\vspace*{-16pt}
			\includegraphics[height=148pt]{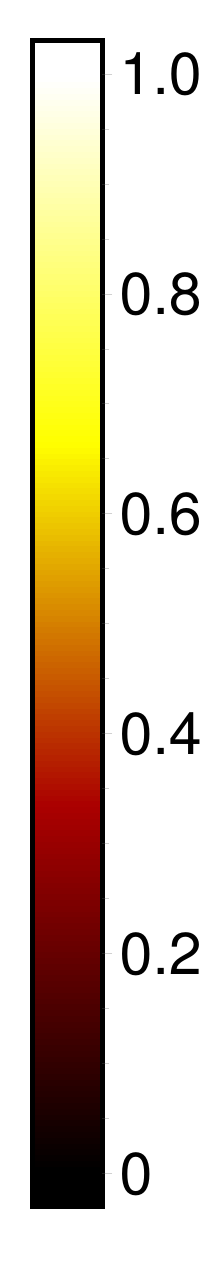}
		\end{minipage}
	\end{center}
	\caption{
	\label{fig:horizon-winding-cusps}
	Left-hand panel: parametric curves of the spherical Kerr horizon (dashed green) and the non-spherical horizon of the local, regular black-hole ($M_\text{exp}$ with $\ell_\text{NP}=0.2556\,m\approx\ell_\text{NP,crit}$), both at spin parameter $a=0.9\,m$.
	Middle panel: Different explicit trajectories (as a function of affine parameter $\lambda$) initiated at image points on the shadow boundary (marked in the right-hand panel) probing symmetric (dashed) and asymmetric (continuous) sections of the spacetime.
	Right-hand panel: The resulting intensity image (normalized to the brightest image point) for inclination $\theta_\text{obs}=\pi/2$. Cusps in the shadow boundary occur whenever trajectories jump between symmetric and asymmetric $\theta$-oscillations (cf.~middle panel).
	}
\end{figure*}

 To proceed in our construction, we require a coordinate invariant characterization of space-time curvature.   
We follow~\cite{1991JMP....32.3135C, 1997GReGr..29..539Z}, provide a comprehensive overview of our construction in an extensive companion paper~\cite{AEAH:2021a} and restrict ourselves to stating the result here: an invariant characterization of local curvature for a Kerr black hole is encoded in
\be
\label{eq:KGR}
K_{\rm GR} = \frac{48 m^2}{\left(r^2+a^2\cos^2(\theta) \right)^3}\;,
\ee
which provides an envelope to the absolute value of the maximum of all independent non-derivative curvature invariants of the Kerr spacetime. Here, we work in units where the Newton coupling $G=1$ and have chosen ingoing Kerr coordinates $u, r, \theta, \phi$. 
We denote the mass and spin parameter by $m$ and $a$, respectively.
To implement principle iv), deviations from Kerr must be functions of $K_{\rm GR} \cdot \ell_{\rm NP}^4$ only, where $\ell_{\rm NP}$ parameterizes the scale of new physics, which is unknown and treated as a free parameter in the following.
\\
To implement principles ii) and iii), we introduce an effective mass function $M(K_{\rm GR})$.
It must interpolate between $M(K_{\rm GR} \rightarrow 0) \rightarrow m$ and $M(K_{\rm GR} \rightarrow \infty) \rightarrow 0$ and must fall off sufficiently fast, when $K_{\rm GR}$ diverges. Specifically, $M(K_{\rm GR}) \sim \left(K_{\rm GR}\cdot \ell_{\rm NP}^4\right)^{-1}$ ensures that curvature invariants are finite and single-valued when 
$\theta \rightarrow \pi/2$ and $r \rightarrow 0$,
in contrast to the Kerr spacetime, which features a ring singularity at this point.\\
Except for these two limits and the requirement of being nowhere singular, $M(K_{\rm GR})$ can, in principle, be chosen freely to construct a model. A specific assumption for the dynamics of the new physics would provide additional constraints on $M(K_{\rm GR})$.
We choose two examples
\bea
\label{eq:Malg}
M_{\rm alg}(K_{\rm GR})&=&\frac{m}{1+\ell_\text{NP}^4K_\text{GR}}\;,
\\
\label{eq:Mexp}
M_{\rm exp}(K_{\rm GR})&=&m\,e^{-(\ell_\text{NP}^4K_\text{GR})^{1/6}}\;.
\eea
$M_\text{alg}$ reduces to a generalization of the Hayward mass function~\cite{Hayward:2005gi} for spherical symmetry. Similarly, $M_\text{exp}$ reduces to the Simpson-Visser mass function~\cite{Simpson:2019mud} in this limit.
For comparison, we choose a non-local model, i.e., one which violates our assumption (iv),
\bea
M_{\rm non-local}(r)
&=&\frac{m}{1+\ell_\text{NP}^4K_\text{GR}(r,\theta=\pi/2)}
\notag\\
&=& \frac{m}{1+\ell_{\rm NP}^4 \frac{48 m^2}{r^6}}\;.
\label{eq:Mnonlocal}
\eea
In the case of a Kerr black hole the mass function is constant, $M(K_\text{GR}) = m=\text{const}$.
\\

\begin{figure*}
	\begin{center}
		\includegraphics[width=0.245\linewidth]{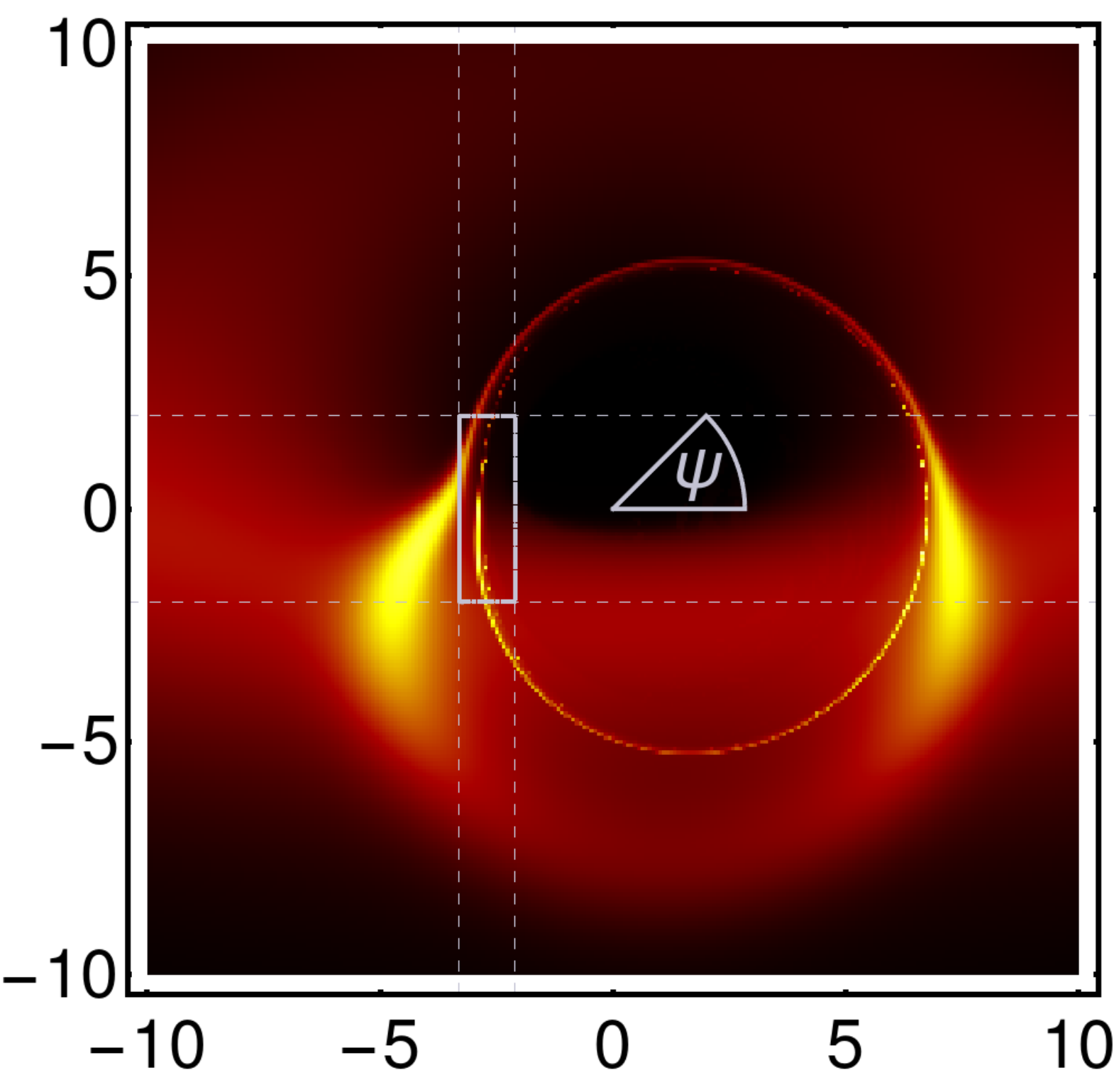}\hfill
		\includegraphics[width=0.245\linewidth]{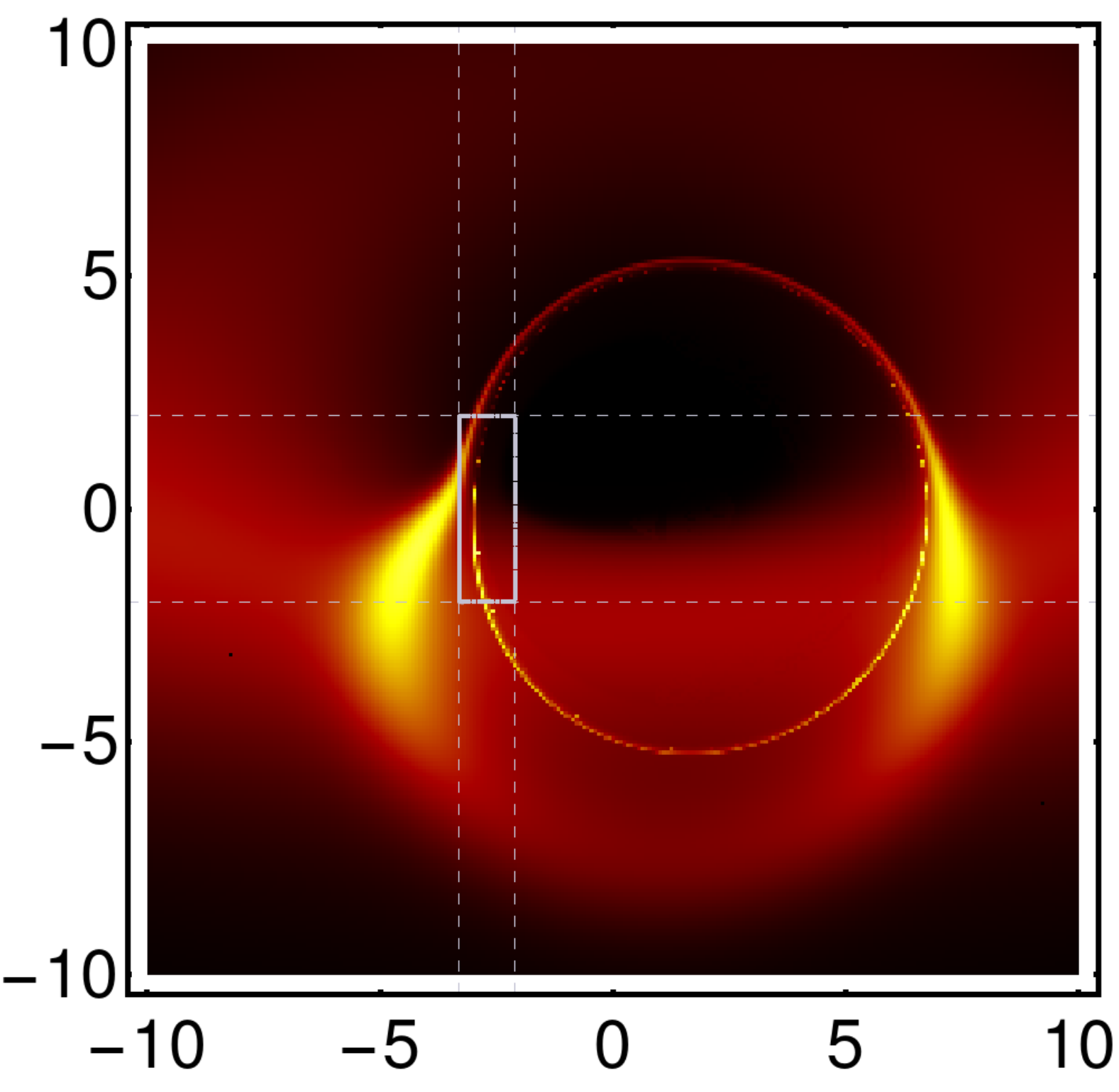}\hfill
		\includegraphics[width=0.245\linewidth]{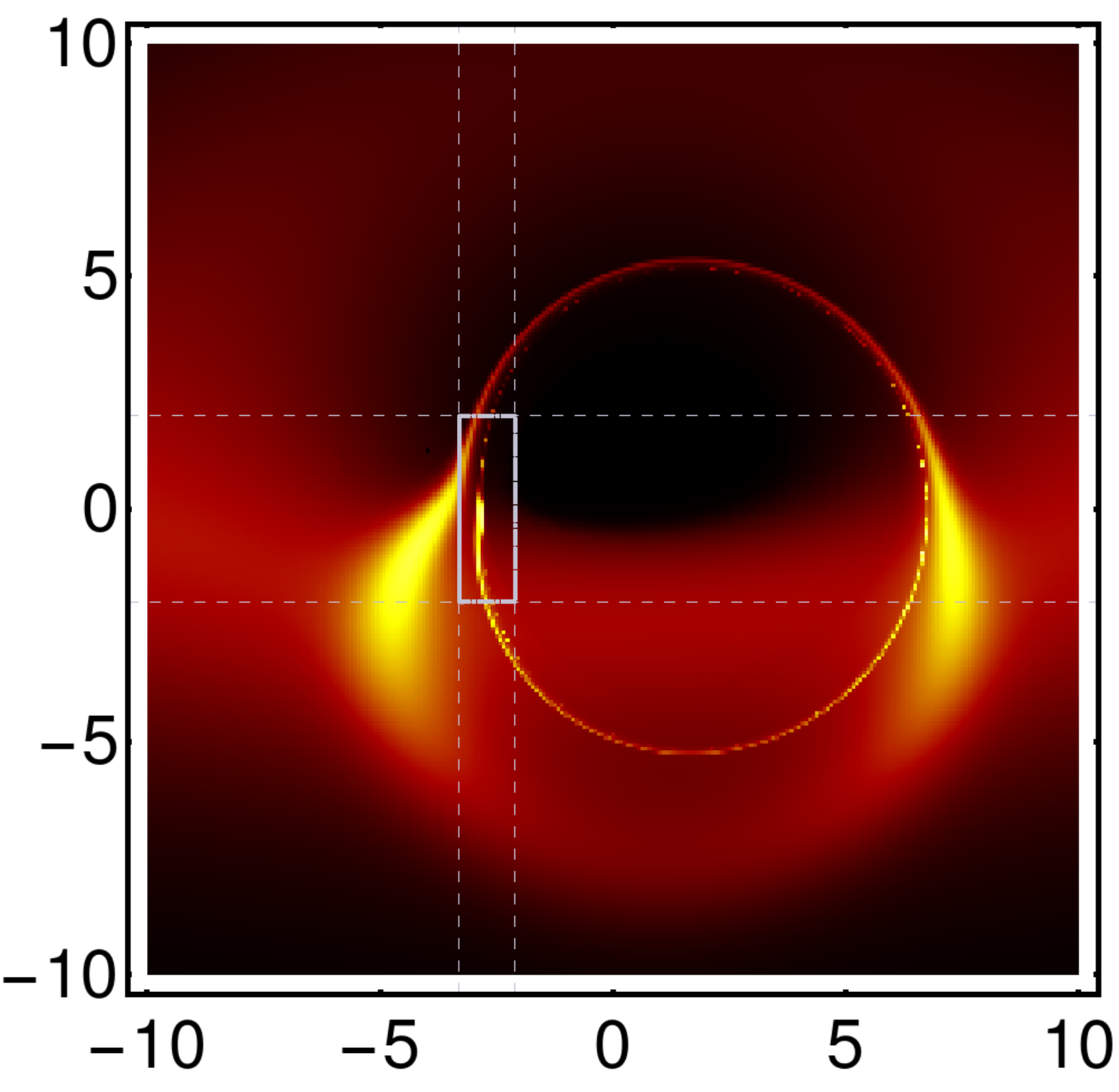}\hfill
		\includegraphics[width=0.245\linewidth]{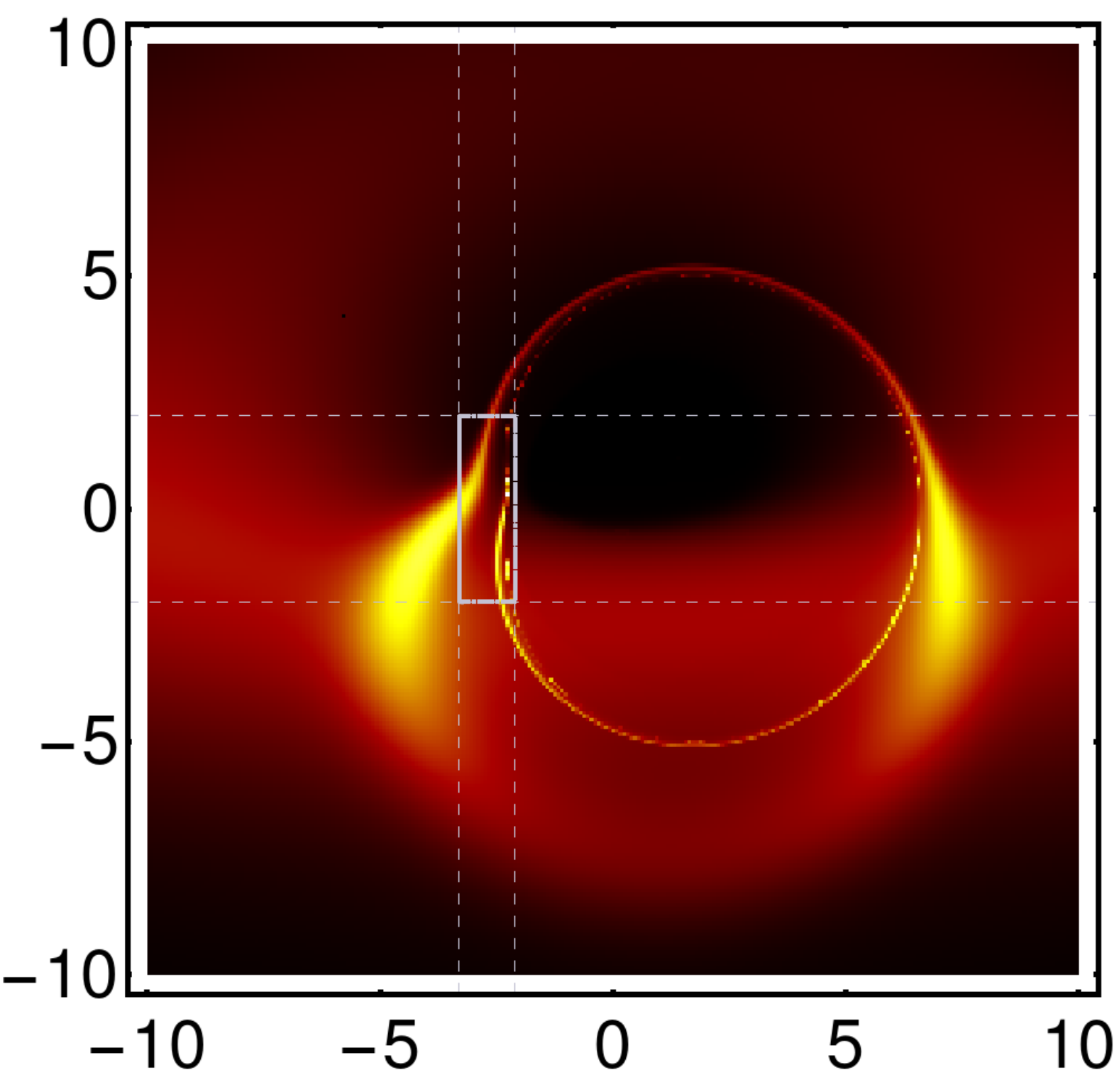}
	\end{center}
	\caption{
	\label{fig:large}
	Relative intensity in the $(x,y)$ image plane, normalized to the brightest image point (of all four images), cf.~color legend in~Fig.~\ref{fig:horizon-winding-cusps}. From left to right: regular black hole that violates the locality principle iv) with $M_\text{non-local}$ with $\ell_\text{NP}=0.2556\,m\approx\ell_\text{NP,crit}$; Kerr black hole; regular black hole that satisfies principle iv) with $M_\text{alg}$ with $\ell_\text{NP}=0.01198\,m\approx\ell_\text{NP,crit}$; regular black hole that satisfies principle iv) with $M_\text{exp}$ with $\ell_\text{NP}=0.2556\,m\approx\ell_\text{NP,crit}$. In all cases, we choose the spin parameter $a=0.9\, m$, inclination $\theta_\text{obs}=8\pi/20$, and disk parameter $h=10\, m$.
	The gray boxes indicate the prograde region in which the deviations from Kerr spacetime are the largest, cf.~Fig.~\ref{fig:details} for detail images.
	}
\end{figure*}

\emph{A dent from local new physics}:
To characterize the black hole spacetime, we determine the location of the event horizon numerically. Intricacies like a deviation of the Killing and the event horizon are spelled out in more detail in a companion paper~\cite{AEAH:2021a}. Here, we focus on two key features of the event horizon, cf.~left-hand panel in Fig.~\ref{fig:horizon-winding-cusps}.

(i) Globally, the horizon shrinks in comparison to a singular Kerr black hole with the same $m$ and $a$. The same effect occurs in many spherically symmetric examples,~see,~e.g.,~\cite{Dymnikova:1992ux,Hayward:2005gi,Platania:2019kyx,Simpson:2019mud,Amir:2016cen,Held:2019xde,Stuchlik2019} and for all three mass functions, including $M_\text{non-local}(r)$. This effect can be traced back to the singularity-resolving new physics which acts as a weakening of gravity and leads to a more compact event horizon. 

(ii) An additional effect is only present in the axisymmetric case: the event horizon features a dent at $\theta=\pi/2$, cf.~left-hand panel in Fig.~\ref{fig:horizon-winding-cusps}, i.e., a minimum of the radius of the horizon as a function of $\theta$. 
The dent is a result of increased compactness in the equatorial plane. This follows from the angular dependence of the mass function $M(K_\text{GR})$ as a result of the classical curvature invariant $K_{\rm GR}$ in Eq.~\eqref{eq:KGR}.

In contrast, the non-local mass function in Eq.~\eqref{eq:Mnonlocal} features no angular dependence and accordingly results in a spherically symmetric event horizon.
More generally, this holds for any mass function which does not depend on $\theta$ and therefore violates our locality principle.
\\

\emph{Impact of astrophysical environments}: 
To make contact with observations, it is critical to study observational signatures in more realistic settings, instead of focusing solely on idealized images features like the shadow boundary. 
Here, we take a first step in this direction. The typical environments of astrophysical black holes are dynamical accretion disks. We consider a simple model of such an environment by adding a simple model of a non-dynamical disk. Further, we neglect absorptivity and make the simplifying assumption that the disk is optically thin. There are strong indications that the accretion disks of supermassive black holes such as M87* and Sgr A* could indeed be optically thin~\cite{Johnson:2015iwg}. Finally, we focus on a single frequency, as is adequate in view of the nearly monochromatic nature of present Event Horizon Telescope observations. In conclusion, we follow~\cite{2020ApJ...897..148G} and model the disk by a density function with variable disk height, i.e.,
\be
	n(r,\theta) = n_0
	\times\exp\left[-\frac{1}{2}\left(\frac{r^2}{100}+h^2\cos^2(\theta)\right)\right],
\ee
where $h$ controls the inverse disk height.
In this setup, radiative transfer along null geodesics is encoded in the following equation for the intensity $I$:
\be
\frac{d}{d\lambda}\left(\frac{I_\nu}{\nu^3}\right) = C\,n\left(x^\mu(\lambda)\right),
\ee
where $\lambda$ is the affine parameter and the radiative transfer equation is evaluated on the photon world line $x^{\mu}(\lambda)$. The frequency is denoted by $\nu$, and $C$ is a constant that, together with the density $n_0$ drops out once we normalize the intensity in the images.
Finally, the radial null geodesics are obtained by numerically solving the geodesic equation.  We mention in passing that the constructed regular spacetimes feature an ergosphere, which is a prerequisite for a jet launching mechanism according to the Blandford-Znajek process~\cite{Blandford:1977ds,McKinney:2005zw}.
\\

\emph{Overall image difference}:
The shadow of spinning Kerr black holes deviates from a spherical shape by about $13\%$ on the prograde side of the image. On this side, light rays are pulled closer to the black hole by frame dragging. The same effect exists for our regular black holes. Thus the images of regular spinning black holes show a similar flattening of the shadow boundary and look largely comparable to the Kerr case, cf.~Fig.~\ref{fig:large}.
Given the finite resolution of the Event Horizon Telescope, our study therefore suggests that the imaged object~\cite{paper1,paper2,paper3,paper4,paper5,paper6} could also be a regular spinning black hole satisfying our four principles.

This motivates the question whether future observations could, in principle, detect imprints of regularity and the locality principle.
Here, we follow a twofold strategy to arrive at a first tentative answer. First, we maximize visible features by using near-critical values of $\ell_\text{NP}$. These are the maximum values such that the compact object features a horizon. Second, we study a range of visibility by testing two mass functions (cf.~Eq.~\eqref{eq:Malg}-\eqref{eq:Mexp}) which result in more or less distinct image features.

In addition, we compare to the case of a non-local mass function (cf.~Eq.~\eqref{eq:Mnonlocal}) to tentatively classify which image features are associated to regularity and which to locality.

From Fig.~\ref{fig:large}, where we compare intensities in the Kerr and the regular cases, the prograde image-side emerges as the interesting image location, which is why we focus on it in the following.
\\

\begin{figure*}
\begin{center}
\begin{tabular}{m{0.02\linewidth}m{0.235\linewidth}m{0.235\linewidth}m{0.235\linewidth}m{0.235\linewidth}}
		$\theta_\text{obs}$ & \hfill$M = m$\hfill${}$ & \hfill$M_\text{non-local}(r)$\hfill${}$ & \hfill$M_\text{alg}(r,\,\theta)$\hfill${}$ & \hfill$M_\text{exp}(r,\,\theta)$\hfill${}$
		\\
		$\frac{\pi}{2}$ &
		\includegraphics[width=\linewidth]{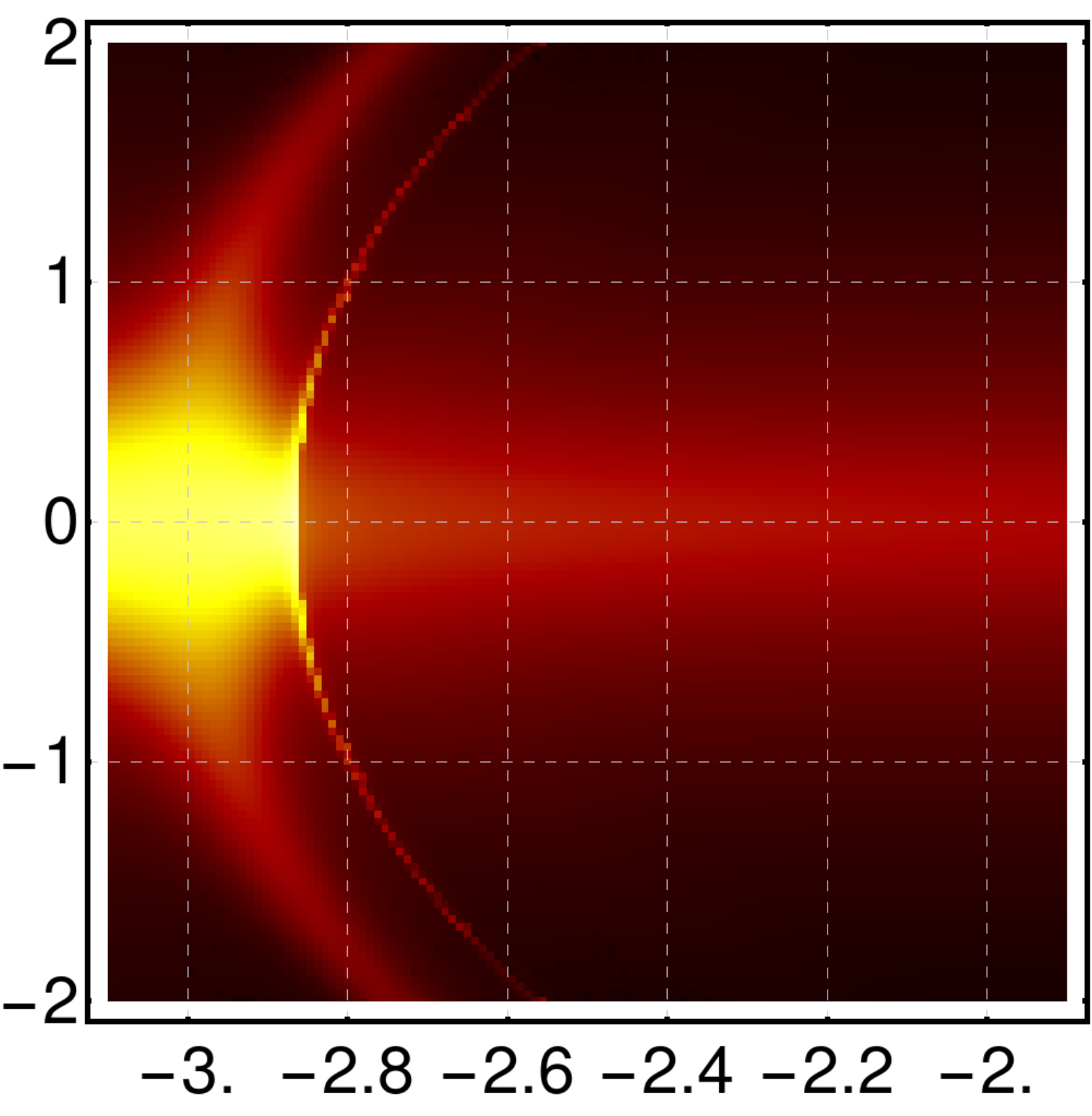} &
		\includegraphics[width=\linewidth]{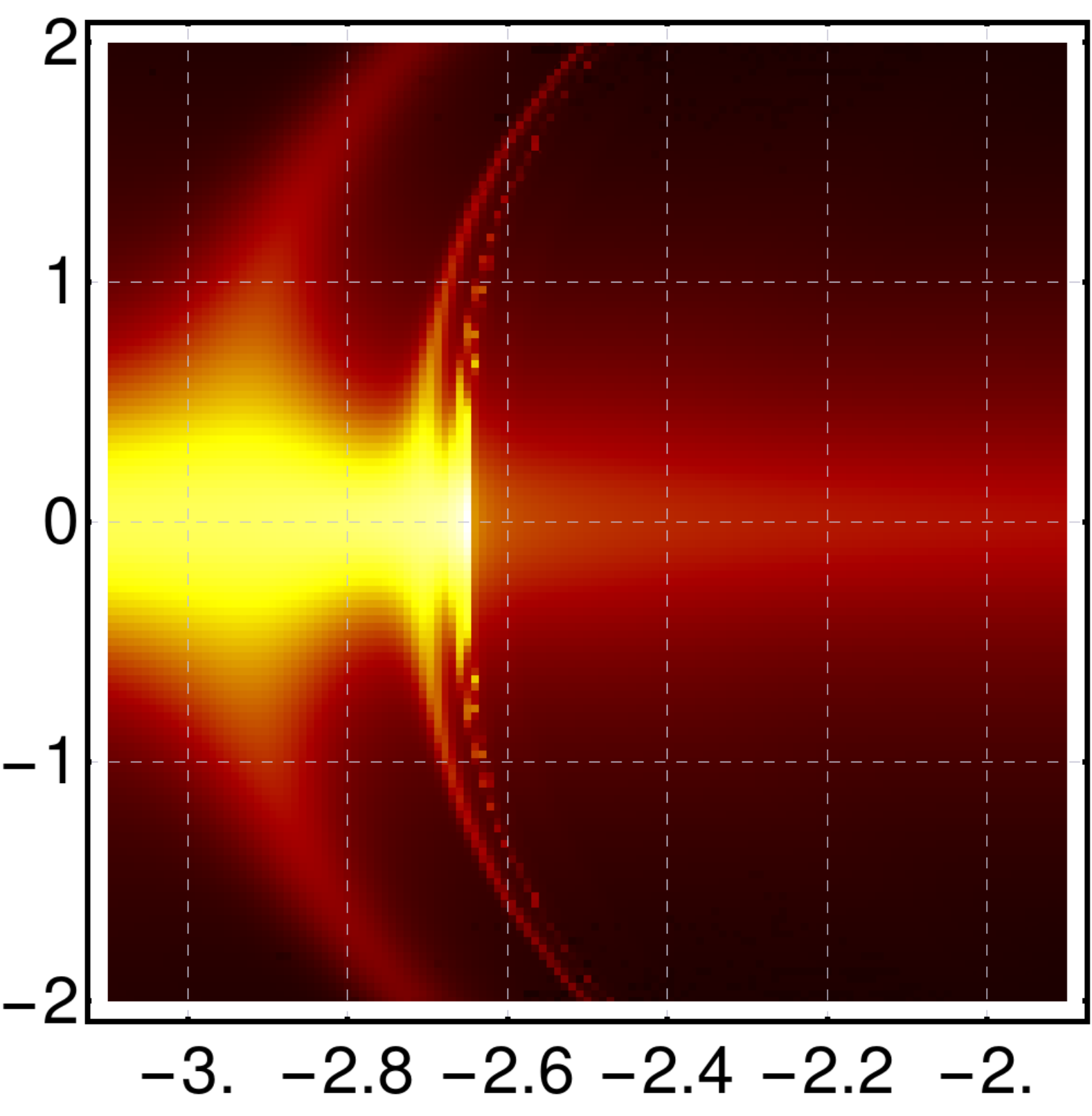} &
		\includegraphics[width=\linewidth]{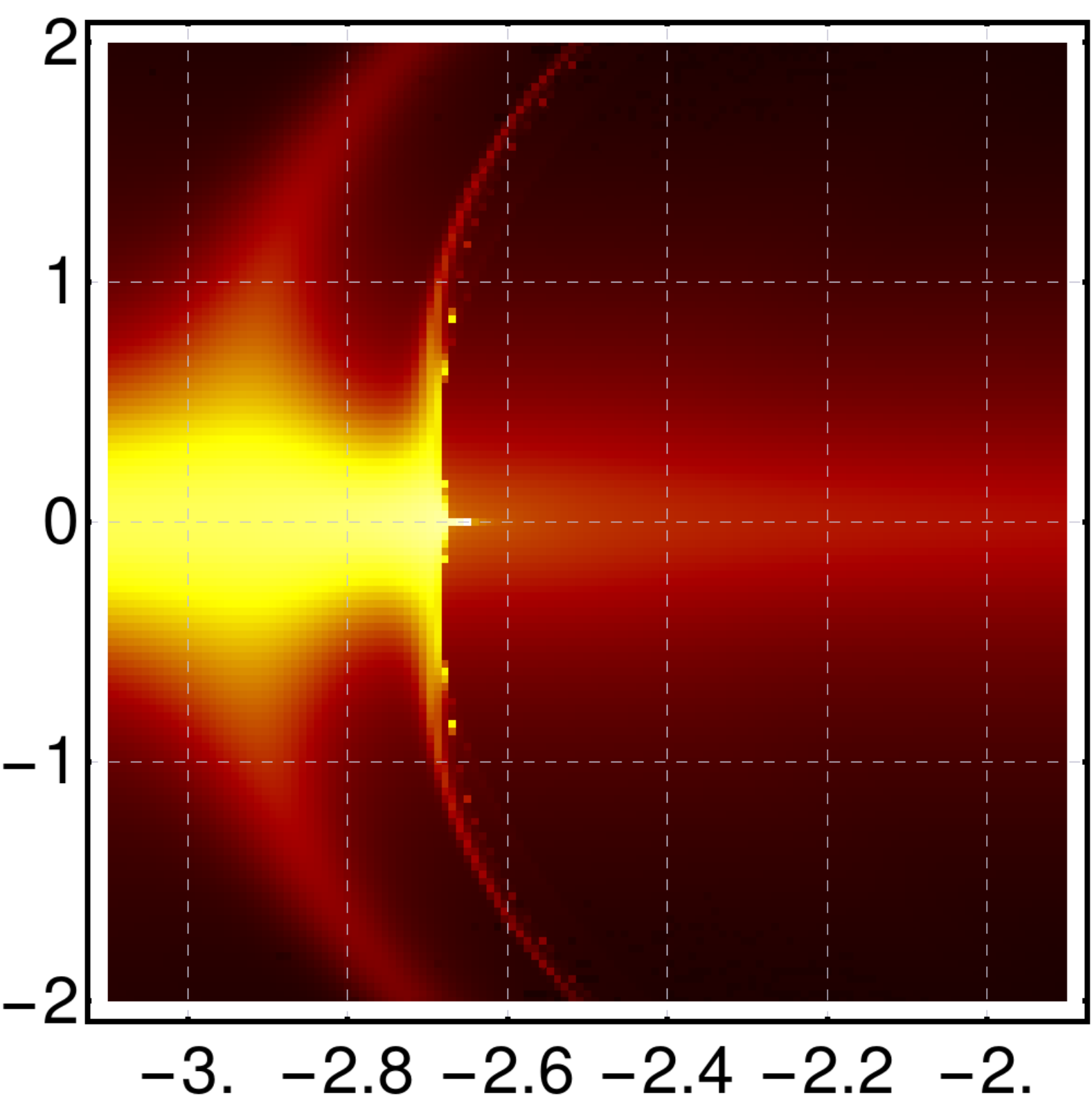} &
		\includegraphics[width=\linewidth]{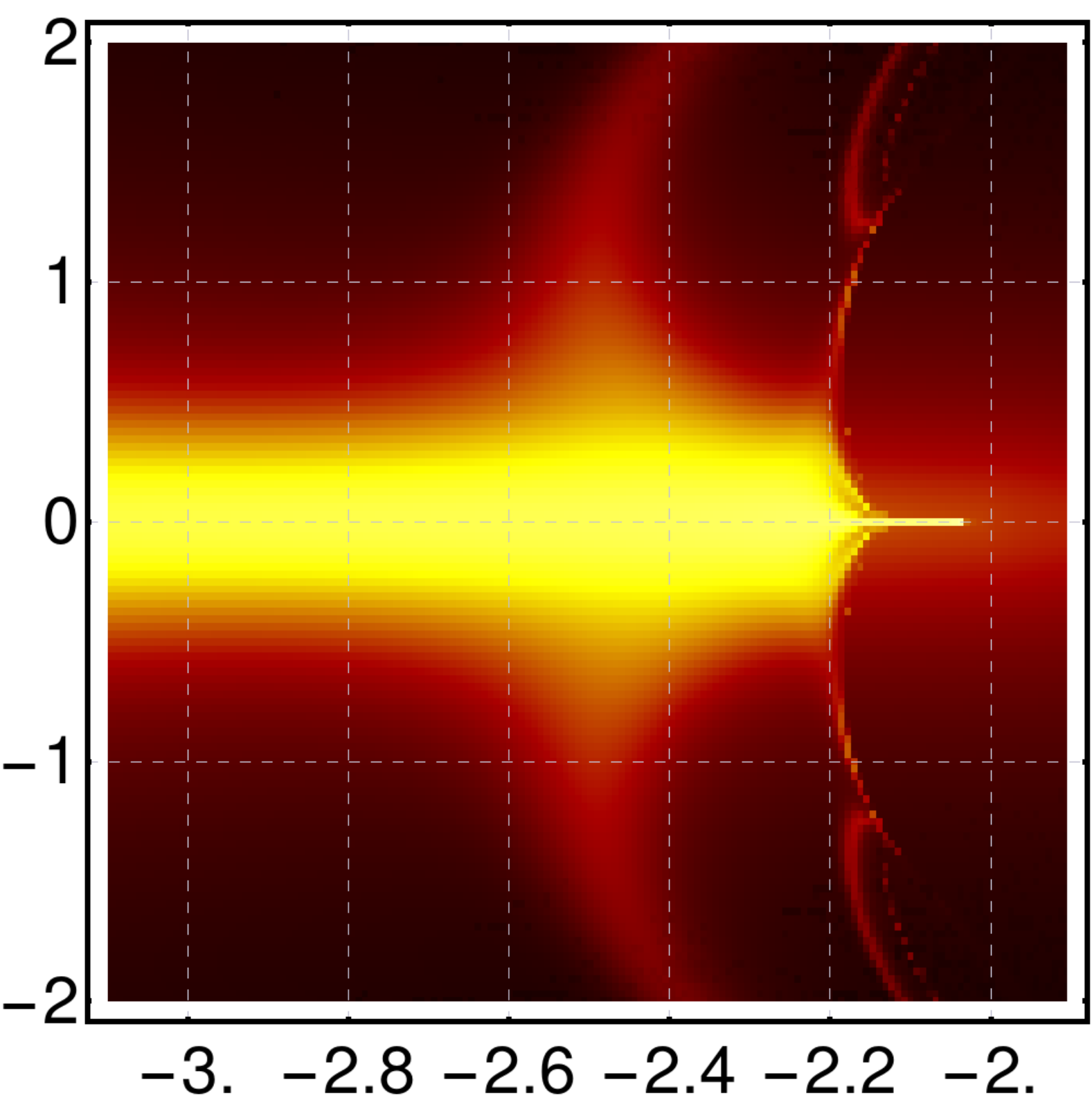}
		\\
		$\frac{9\pi}{20}$ &
		\includegraphics[width=\linewidth]{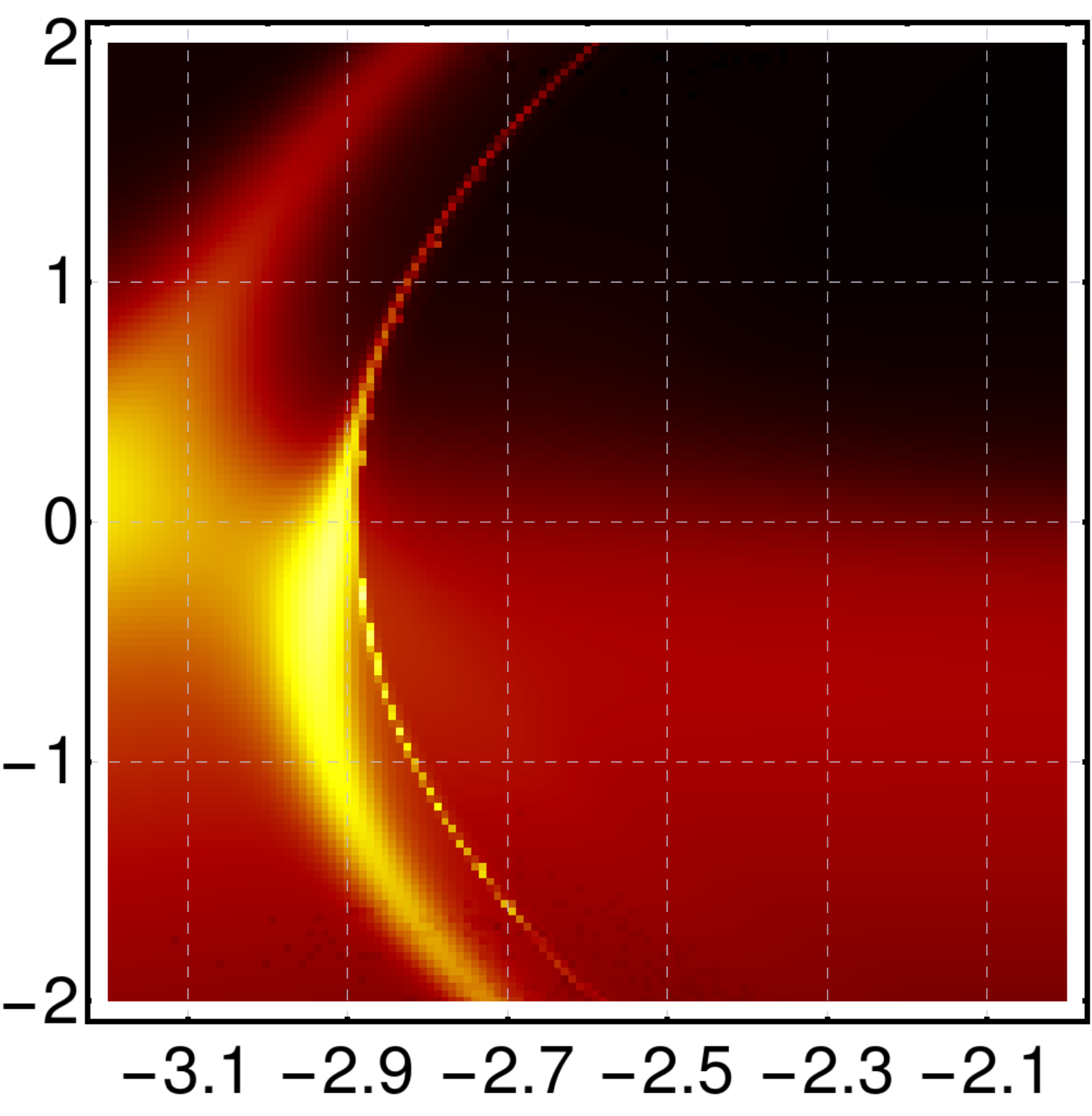} &
		\includegraphics[width=\linewidth]{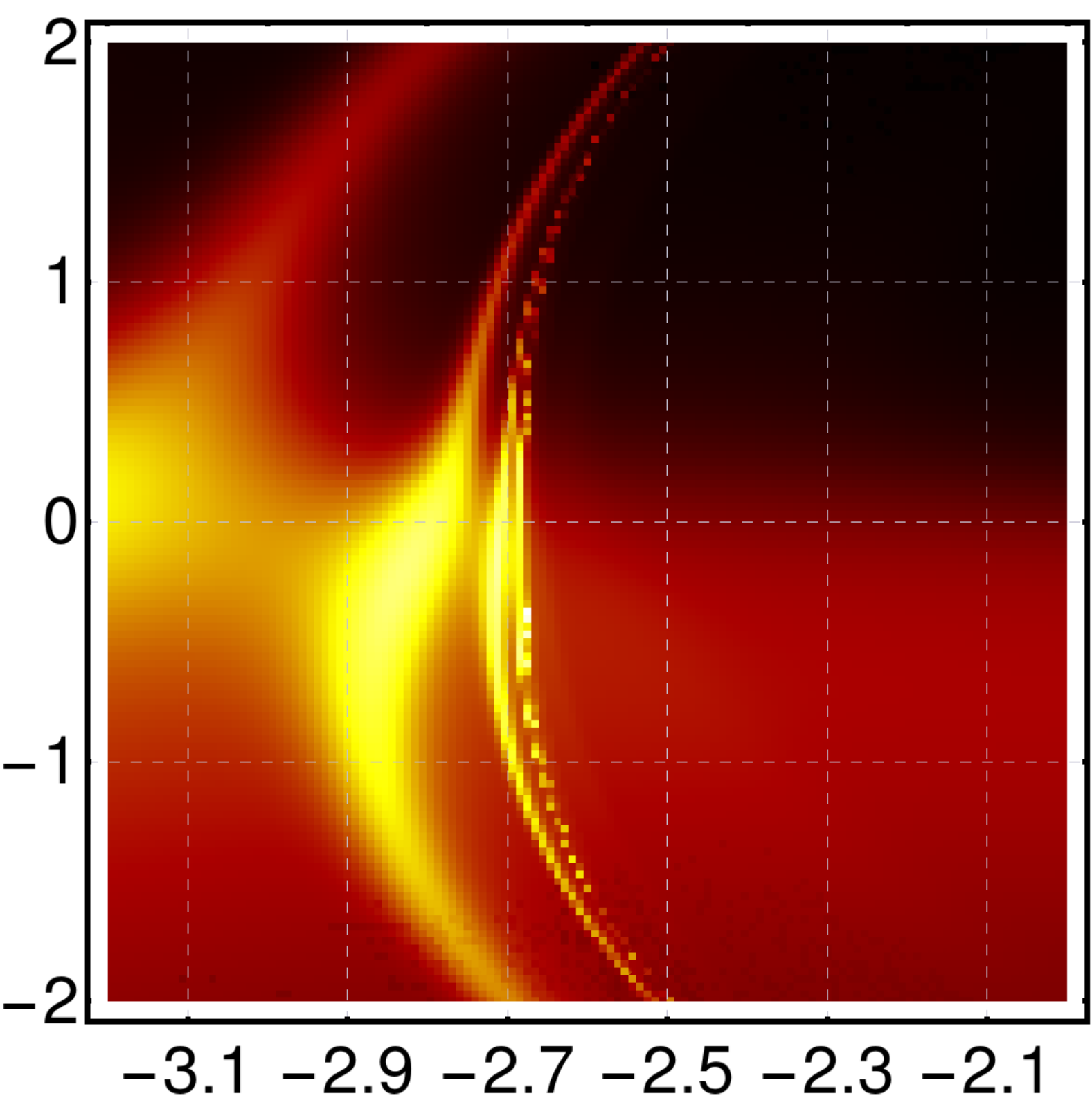} &
		\includegraphics[width=\linewidth]{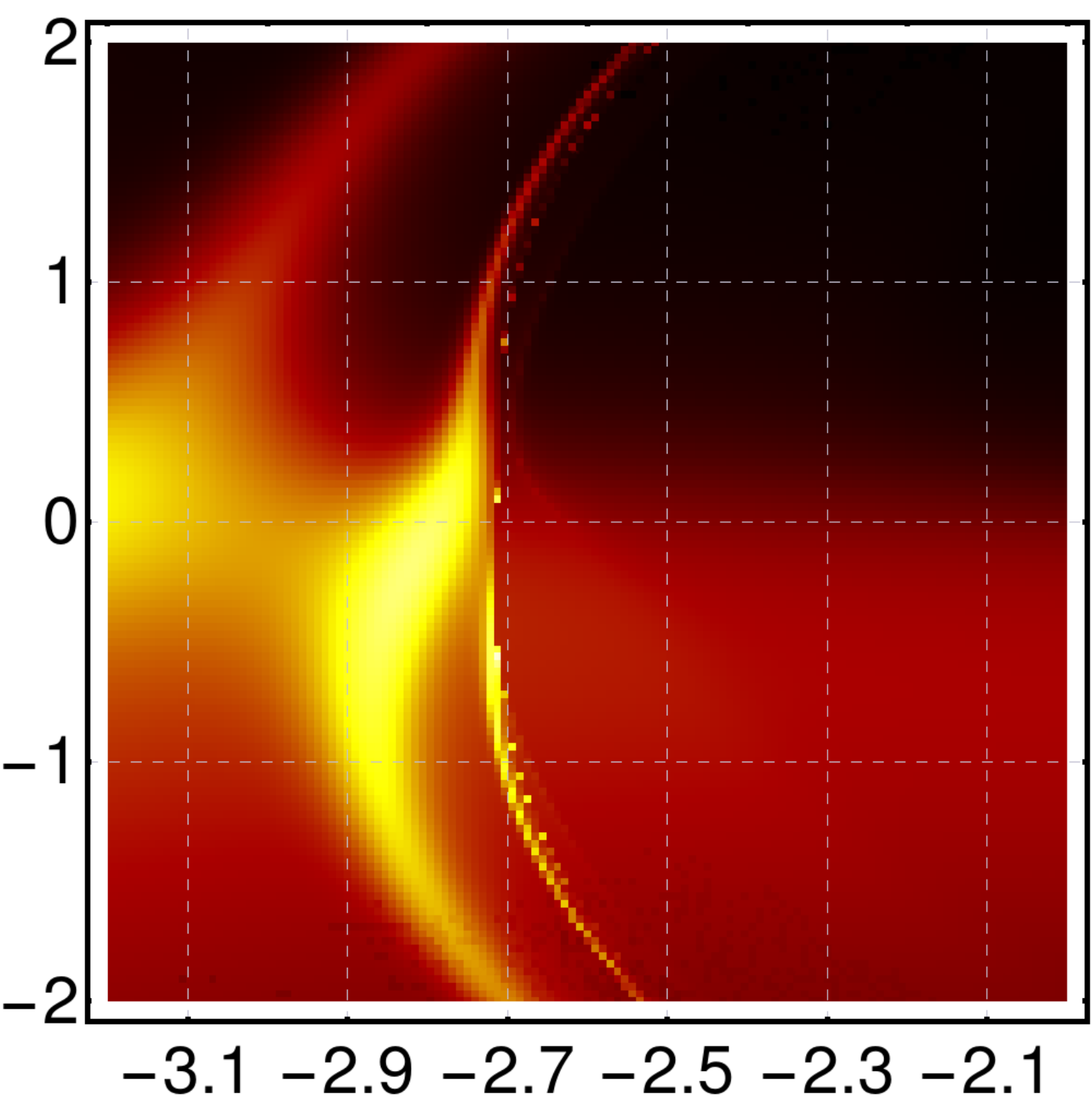} &
		\includegraphics[width=\linewidth]{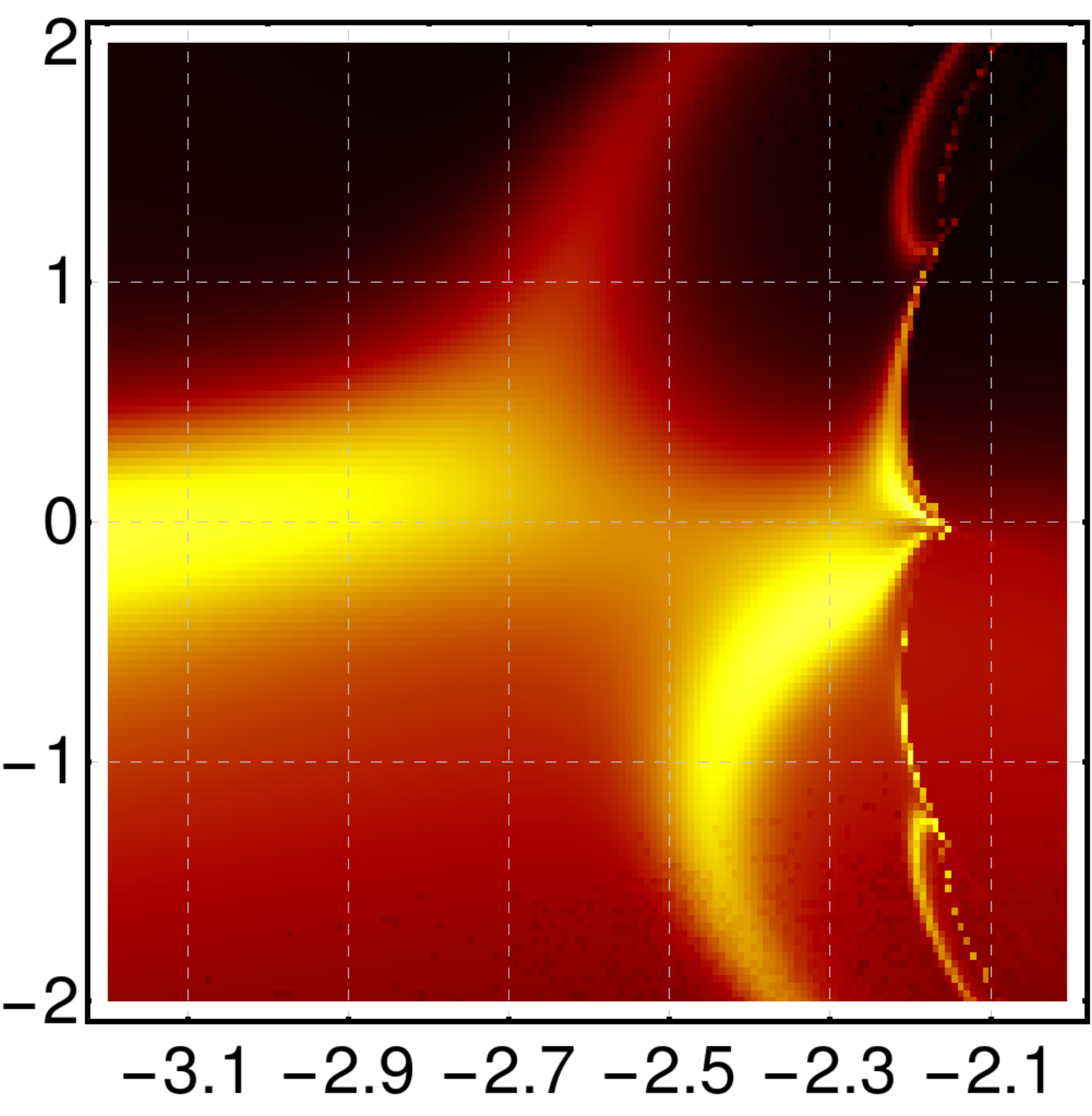}
		\\
		$\frac{8\pi}{20}$ &
		\includegraphics[width=\linewidth]{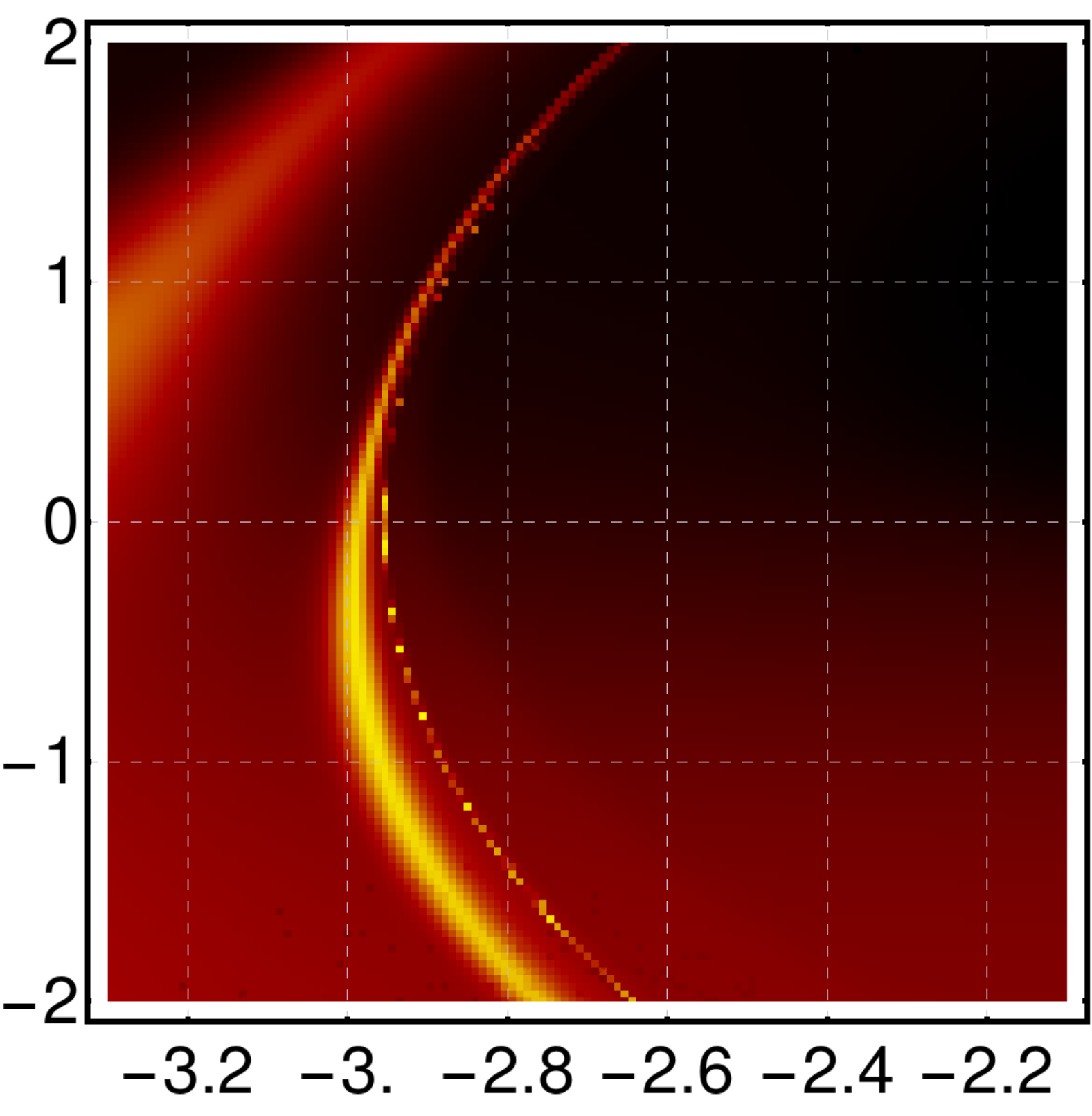} &
		\includegraphics[width=\linewidth]{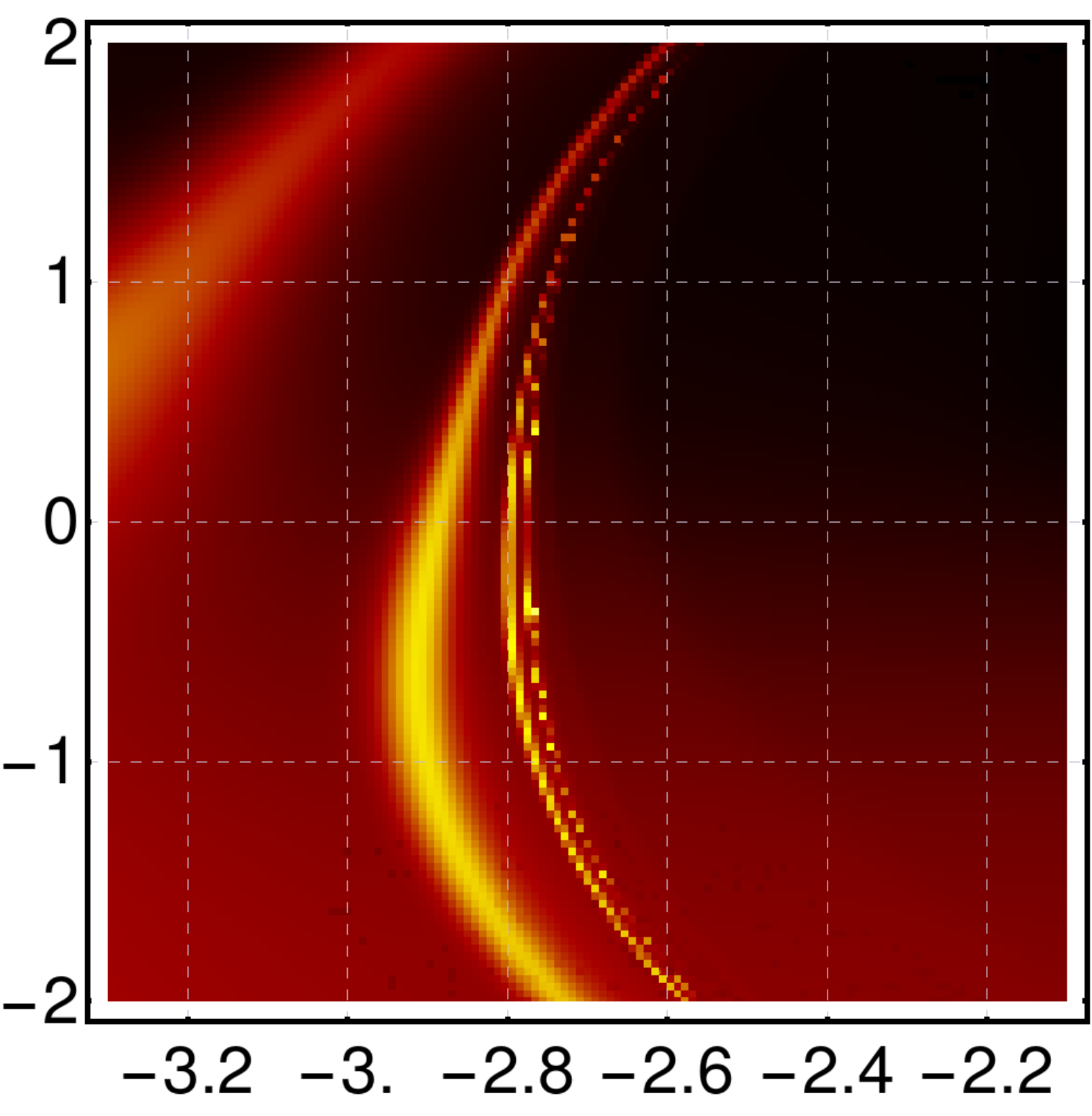} &
		\includegraphics[width=\linewidth]{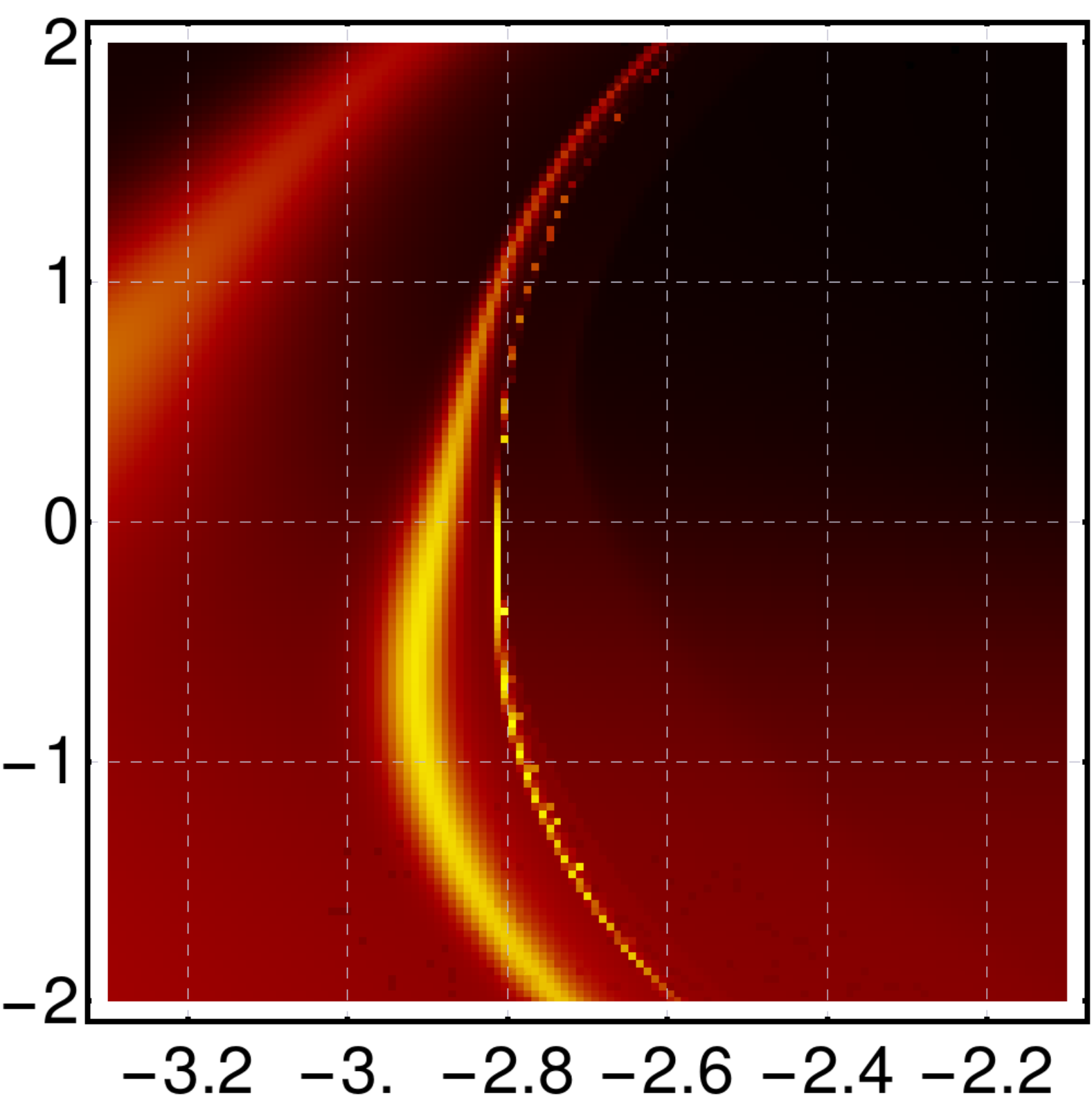} &
		\includegraphics[width=\linewidth]{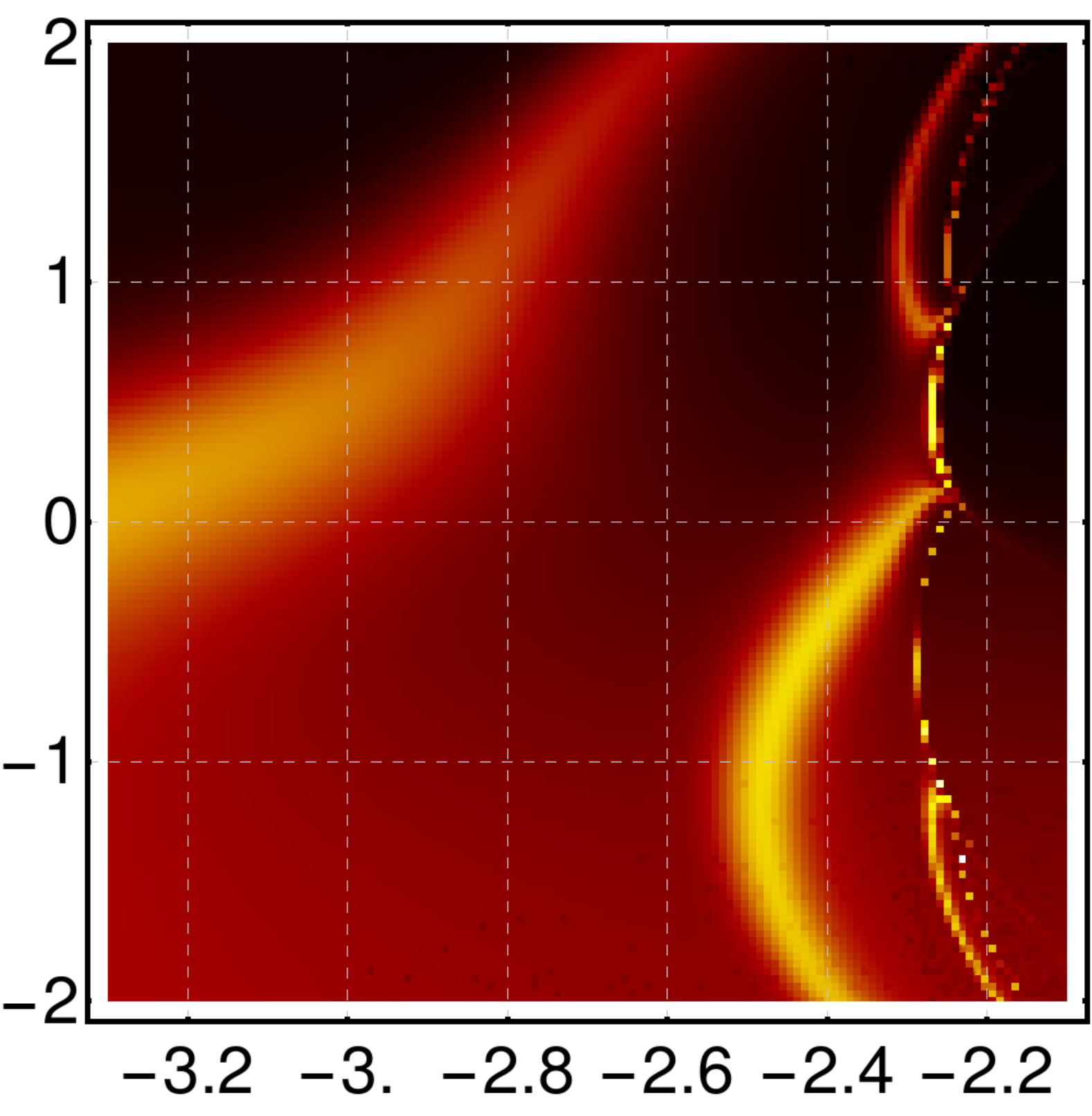}
		\\
		$\frac{17\pi}{180}$ &
		\includegraphics[width=\linewidth]{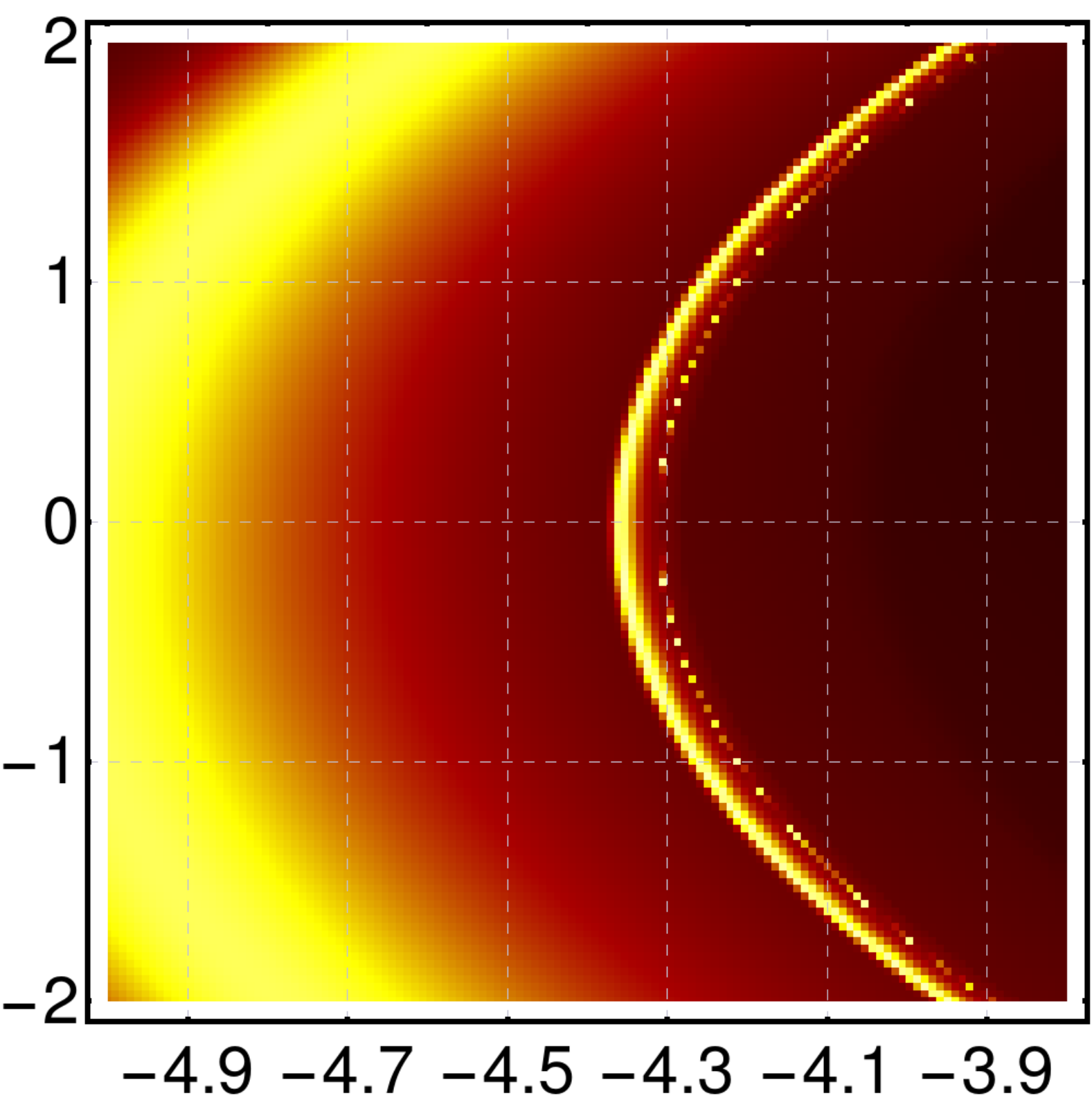} &
		\includegraphics[width=\linewidth]{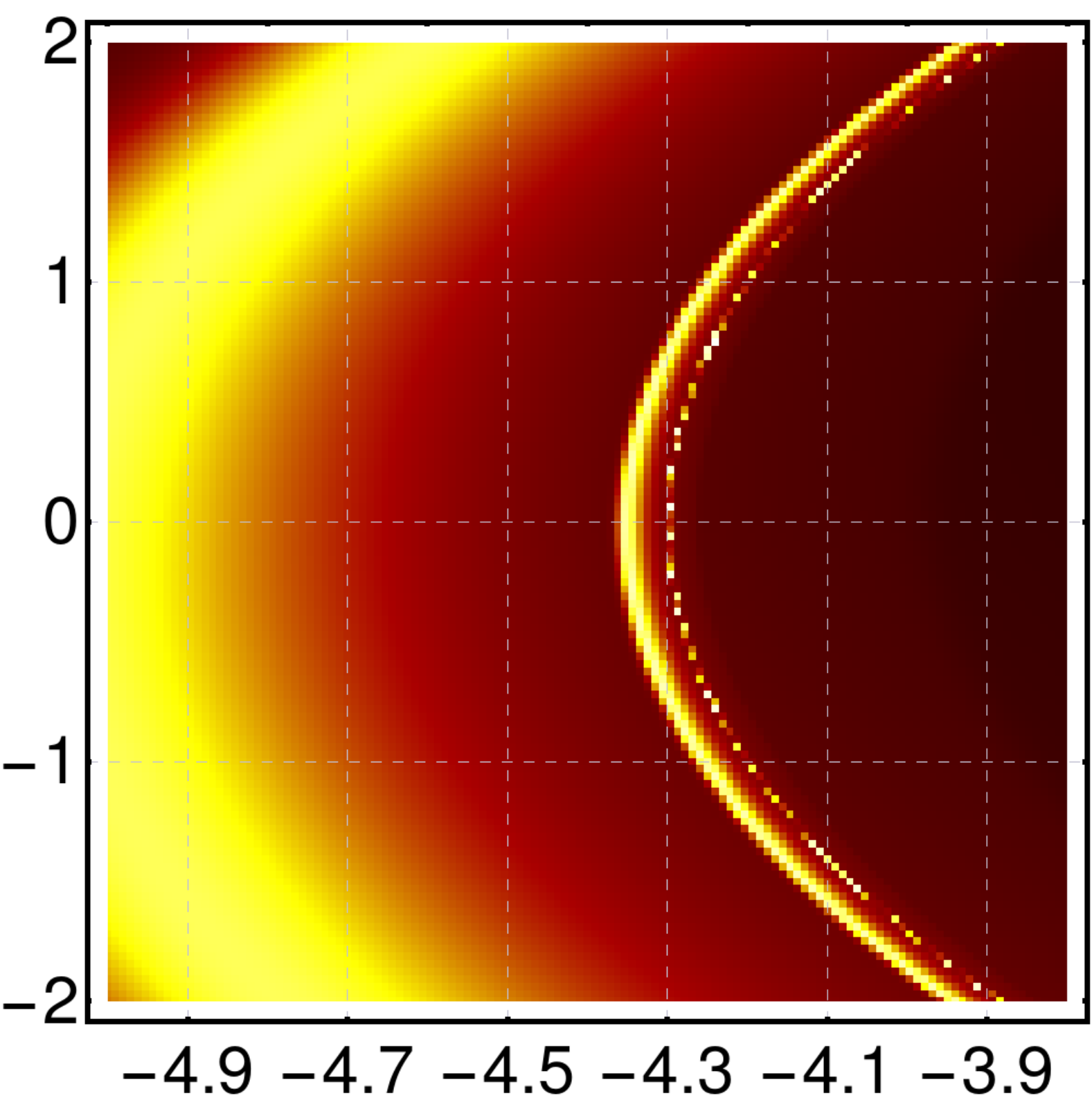} &
		\includegraphics[width=\linewidth]{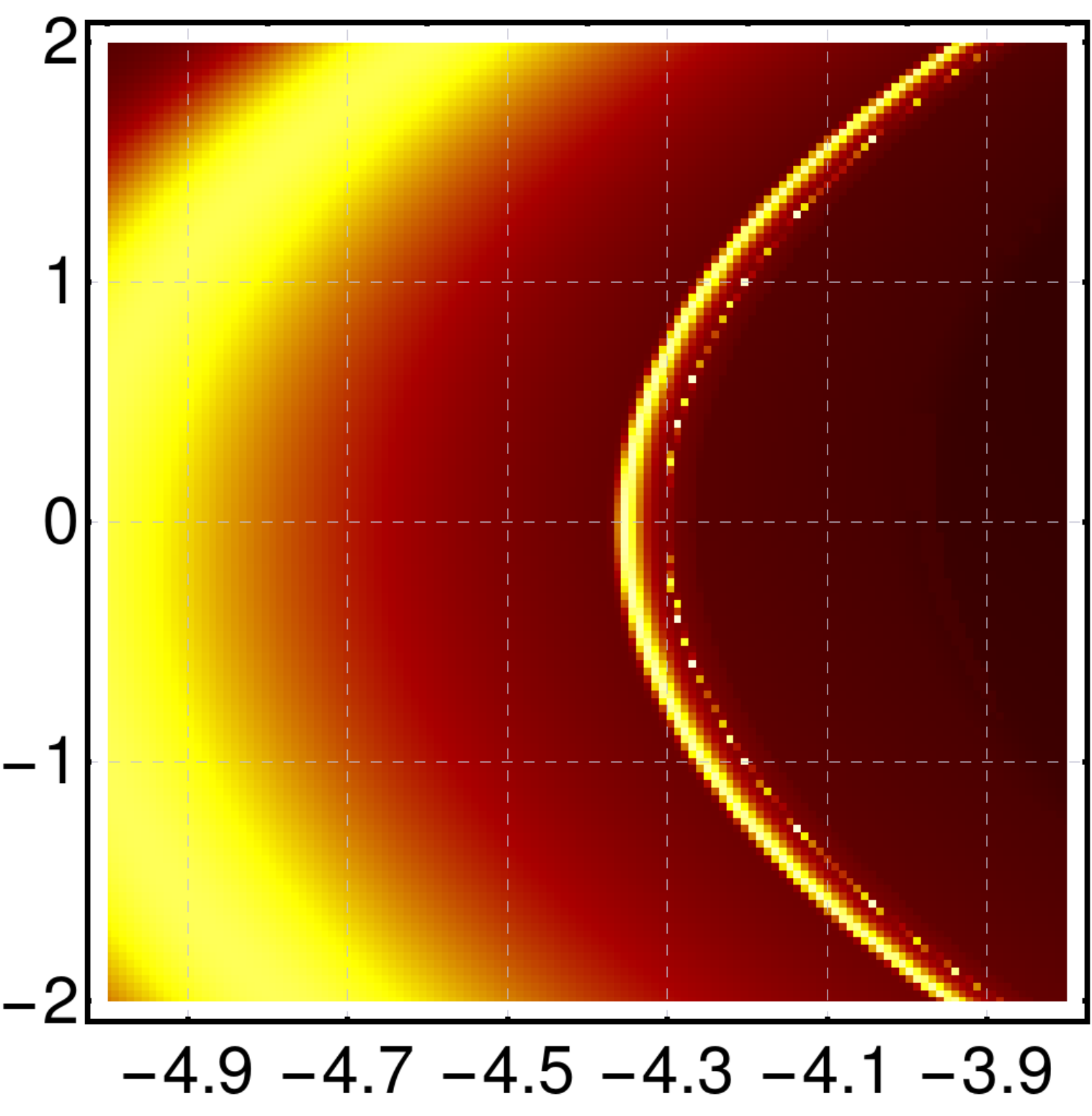} &
		\includegraphics[width=\linewidth]{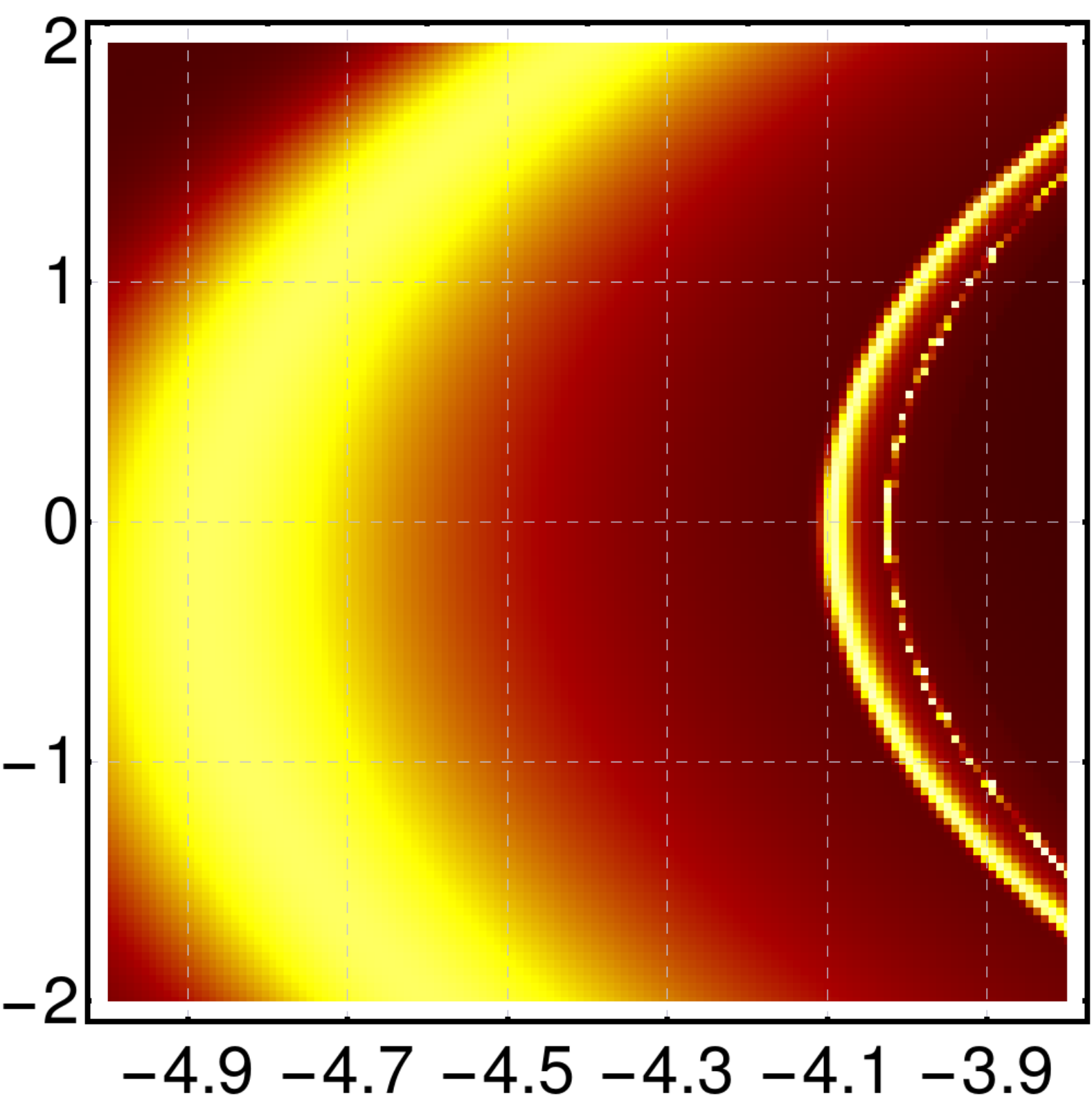}
	\end{tabular}
\end{center}
	\caption{
	\label{fig:details}
	Detailed intensity images of the prograde side of spinning black holes ($a=0.9\,m$) in the $(x,y)$ image plane. In each row, we normalize the intensity to the brightest image point (of all four images), cf.~color legend in~Fig.~\ref{fig:horizon-winding-cusps}. Columns from left to right correspond to mass functions $M=m$ (Kerr), $M_\text{non-local}(r)$ (cf.~Eq.~\eqref{eq:Mnonlocal}), $M_\text{alg}(r,\,\theta)$ (cf.~Eq.~\eqref{eq:Malg}), and $M_\text{exp}(r,\,\theta)$ (cf.~Eq.~\eqref{eq:Mexp}), respectively. Apart from the Kerr case, the black-hole spacetimes are regular everywhere and, as in Fig.~\ref{fig:large}, $\ell_\text{NP}$ is chosen close (i.e., with 4-digit precision) to the critical value $\ell_\text{NP,crit}$ for which, in the respective spherically-symmetric case, the event horizon disappears. The latter maximizes the potential new-physics effect in the presence of an event horizon. The different rows depict different inclinations between the vector pointing from the black hole to the observer and the black hole's spin axis, i.e., $\theta_\text{obs}=\frac{\pi}{2},\,\frac{9\pi}{20},\,\frac{8\pi}{20},\,\frac{17\pi}{180}$ from top to bottom.
	}
\end{figure*}

\emph{Shift of the prograde shadow boundary}:
The shadow is more compact for regular than for Kerr black holes due to the more compact horizon. The effect is significantly larger on the prograde side, where frame dragging enables light rays to probe the geometry closer to the more compact horizon. The resulting appreciable shift of the prograde shadow boundary is not tied to the locality principle and also occurs for non-local mass functions $M(r)$,  cf.~Fig.~\ref{fig:large}, see also, e.g.,~\cite{Amir:2016cen,Abdujabbarov:2016hnw,Tsukamoto:2017fxq,Wang:2018prk,Dymnikova:2019vuz,Contreras:2019cmf,Shaikh:2019fpu,Kumar:2020owy, Ghosh:2020ece,Liu:2020ola}.

This increase in compactness is not degenerate with the flattening effect of increased spin parameter $a$ in the Kerr case. The overall shape of the shadow boundary cannot be reproduced by any Kerr geometry, see~\cite{AEAH:2021a}.
\\

\emph{Relative stretching}:
Gravitational lensing in the vicinity of the black-hole horizon results in multiple lensed images of the disk. Corresponding image features can vary significantly with inclination but are visible at all inclinations, cf.~Fig.~\ref{fig:details}. This includes the M87* case, where the spin axis nearly points towards the observer, i.e., $\theta_\text{obs}\approx 17\pi/180$,~\cite{paper5}. The distance between these features is stretched out in the case of regular black holes. The intensity peaks furthest away from the shadow boundary lie at similar locations in the image plane in the singular and regular case, cf.~Fig.~\ref{fig:large}. As null geodesics move towards the light sphere, the compactness of the probed geometry increases. Therefore, the distance to further inwards lying intensity maxima is stretched out, in comparison to the Kerr case.

The amount of stretching is a function of the radial and angular dependence of the mass function $M(r,\,\theta)$. In particular, $M_\text{exp}$ leads to a larger effect in comparison to $M_\text{alg}$. In both cases, the amount of stretching can vary significantly with image angle $\psi$ (indicated in the left-most panel in Fig.~\ref{fig:large})
because the mass function also depends on $\theta$. Therefore the increase in compactness has an angular ($\psi$) dependence.
In contrast, for a non-local mass function, the effect is more homogeneous across different image angles because the increase in compactness only depends on the radius.

Away from the shadow boundary, image features depend on the geometry as well as on the emission structure~\cite{Broderick:2005jj,Broderick:2016ewk,Tiede:2020jgo}. A first study of the impact of varying disk parameter $h$ in our specific disk model is conducted in~\cite{AEAH:2021a} and supports that, in the given model, image features persist.
\\

\emph{Cusps from the locality principle}:
The overall shift as well as the stretching out of image features distinguish spinning regular black holes from Kerr black holes, but are present for mass functions that either satisfy or violate the locality principle. 
This motivates us to search for potential distinguishing features between the two classes of regular mass functions. 
A characteristic difference between the two classes of geometries is the dent in the horizon, exclusively present for mass functions satisfying the locality principle.
Indeed, we find that the latter can result in characteristic cusp-like features in the shadow images, cf.~Fig.~\ref{fig:details}.
More specifically, the cusps are discontinuities in the shadow boundary. Just like shift and relative stretching, these features are enhanced for $M_\text{exp}$ compared to $M_\text{alg}$. Moreover, they are most distinct when the regular black holes are viewed face-on ($\theta_\text{obs}=\pi/2$).

To understand the origin of the cusps, one needs to know that geodesics making up the shadow boundary cover bounded ranges in $\theta$ while orbiting the black hole~\cite{Vazquez:2003zm,Gralla:2019ceu,Himwich:2020msm}. The range between $\theta_{\rm min}(\psi)$ and $\theta_{\rm max}(\psi)$ depends on the image angle $\psi$. When the range of $\theta$ changes, the corresponding radius of the horizon differs, cf.~left panel in Fig~\ref{fig:horizon-winding-cusps}. Thus, null geodesics that arrive at distinct $\psi$ effectively probe different sections of the near-horizon geometry. For instance, trajectories on two sides of a cusp probe symmetric and asymmetric sections of the horizon geometry, respectively, cf.~middle panel in Fig.~\ref{fig:horizon-winding-cusps}.
Discontinuous jumps in $\theta_{\rm min/max}(\psi)$ result in discontinuities in the shadow boundary at $\psi$.

In particular, $\psi=\pi$ (or $\psi = 0$ in case of flipped spin parameter $a$) is associated to a particularly pronounced discontinuity at $\theta_{\rm obs} \approx \pi/2$, cf.~right-hand panel in Fig.~\ref{fig:horizon-winding-cusps}. In this case, the dent in the event horizon is mirrored by a corresponding dent in the image, see also~\cite{Held:2019xde}.

In contrast, for a spherically symmetric horizon, any section of the near-horizon geometry contains the same information on a constant horizon radius. Therefore changes in $\theta_\text{min/max}$ as a function of $\psi$ do not lead to discontinuities in the shadow boundary. This holds for singular Kerr black holes. Similarly, we expect that any non-local mass function that is also associated with a spherically symmetric event horizon results in a continuous shadow boundary as exemplified by $M_\text{non-local}$.\\

\emph{Broken reflection symmetry from the locality principle}:
From the images in Fig.~\ref{fig:details}, we observe that the lack of reflection symmetry about the $y=0$ axis generally characterizes images which are not face on, i.e.,  for which $\theta_{\rm obs}\neq \pi/2$. For a Kerr black hole, the asymmetry decreases towards and vanishes at the shadow boundary. The same observation holds for the non-local mass function, but not in the case of a local mass function. For $M_{\rm alg}$ and $M_{\rm exp}$, reflection symmetry of the shadow boundary is broken. This is another consequence of an event horizon that breaks spherical symmetry. In fact,
intuition based on geometric optics would suggest that when one views a non-spherically symmetric object, the resulting image depends on whether the object is tilted towards or away from the observer. In particular,  even if the object has a reflection symmetry about its equatorial plane, images at nontrivial inclination break this symmetry.
This intuition describes the impact of the inclination on the shape of the shadow boundary.

Again, as for the cusp-like features, the asymmetry follows from the dent in the event horizon and therefore from our locality assumption.
\\

\emph{Outlook: Universality classes of new-physics principles.} Our results suggest that regular black holes could be characterized by a set of physical principles that result in qualitatively distinct image features, constituting universality classes for black-hole shadows. A regularity principle results in a more compact shadow, a shifted prograde shadow boundary, and relative stretching of image features. All representatives of this principle that we have explored as well as others in the literature~\cite{Amir:2016cen,Abdujabbarov:2016hnw,Tsukamoto:2017fxq,Wang:2018prk,Dymnikova:2019vuz,Contreras:2019cmf,Shaikh:2019fpu,Kumar:2020owy, Ghosh:2020ece,Liu:2020ola} exhibit these characteristics. Moreover, the locality principle could characterize a more constrained universality class with added image features, namely cusps and broken reflection symmetry. 

We tentatively conjecture that a distinction of mass functions within a given universality class is rather difficult to achieve given finite EHT resolution capabilities,  see however~\cite{Blackburn:2019bly,2019BAAS...51g.176P}. However, a distinction between such universality classes might potentially be achievable in the case of large $\ell_\text{NP}$. This would constitute physical insight into principles underlying a more fundamental theory of gravity than GR.

These findings motivate a number of upgrades of our study, most importantly a fully dynamical accretion disk with more complete modelling of absorptivity and emissivity, as in~\cite{Porth:2019wxk}.\\

\emph{Acknowledgements:} We gratefully acknowledge discussions with R.~Gold.  A.~E.~is supported by a research grant (29405) from VILLUM fonden. 
A.~H.~is supported by a Royal Society Newton fellowship [NIF/R1/191008].

\bibliography{References}

\end{document}